\documentclass[%
 reprint,
amsmath,amssymb,
aps,
]{revtex4-2}

\usepackage{graphicx}
\usepackage{dcolumn}
\usepackage{bm}
\usepackage{physics}
\usepackage{verbatim}
\usepackage[small,tight,FIGTOPCAP]{subfigure}
\usepackage{amsmath}
\usepackage{amsfonts}
\usepackage{amssymb}
\usepackage[]{mathrsfs}
\usepackage{xcolor}
\usepackage{bbm}
\usepackage [english]{babel}

\usepackage{graphicx}
\usepackage{dcolumn}
\usepackage{bm}
\usepackage{physics}
\usepackage{bbm}
\usepackage{hyperref}
\usepackage{color}
\usepackage{hyperref}
\usepackage{cancel}

\begin{document}

\preprint{APS/123-QED}

\title{Observing super-quantum correlations across the exceptional point in a single, two-level trapped ion}%

\author{A. Quinn$^{1}$}\thanks{These two authors contributed equally.} 
\author{J. Metzner$^{1}$}\thanks{These two authors contributed equally.}
\author{J.E. Muldoon$^2$, I.D. Moore$^1$, S. Brudney$^1$, S. Das$^3$}
\author{D.T.C. Allcock$^1$}\thanks{E-mail: dallcock@uoregon.edu, yojoglek@iu.edu.}
\author{Y.N. Joglekar$^2$}\thanks{E-mail: dallcock@uoregon.edu, yojoglek@iu.edu.} 

\affiliation{$^1$ Department of Physics, University of Oregon, Eugene, OR, USA}
\affiliation{$^2$ Department of Physics, Indiana University Purdue University Indianapolis (IUPUI), Indianapolis, IN, USA}
\affiliation{$^3$ Department of Physical Sciences, Indian Institute of Science Education and Research (IISER) Kolkata, Mohanpur 741246, West Bengal, India}

\date{\today}

\begin{abstract}
Quantum theory provides rules governing much of the microscopic world, and among its counter-intuitive consequences are correlations that exceed the bounds from local, classical theories. In two-level quantum systems - qubits - unitary dynamics theoretically limit these spatiotemporal quantum correlations, called Bell/Clauser-Horn-Shimony-Holt~\cite{Bell2004,chsh1969} or Leggett-Garg inequalities~\cite{Leggett1985,Emary2013}, to $2\sqrt{2}$ or 1.5  respectively. Experiments with state-of-the-art qubits have approached the spatial, Bell~\cite{Hensen2015,2018} and temporal, Leggett-Garg~\cite{PalaciosLaloy2010,Athalye2011,Waldherr2011,Goggin2011,Knee2016} quantum correlation bounds. Here, using a dissipative, trapped $^{40}$Ca$^+$ ion governed by a two-level, non-Hermitian Hamiltonian, we observe temporal correlation values up to $1.703(4)$ for the Leggett-Garg parameter $K_3$, clearly exceeding the hitherto inviolable L\"{u}der's  bound of 1.5. These excesses occur across the exceptional point of the parity-time symmetric Hamiltonian responsible for the qubit's non-unitary, coherent dynamics. Distinct evolution speeds for antipodal qubit states, which violate the unified (Mendelstam-Tamm or Margolus-Levitin) bound $\tau_{\textrm{QSL}}$ for the transit time based on quantum speed limit, result in the super-quantum $K_3$ values observed over a wide parameter range. Our results demonstrate that post-selected, coherent dynamics of non-Hermitian Hamiltonians pave the way for enhanced quantum correlations that exceed protocols based on unitary or dissipative dynamics. 
\end{abstract}

\maketitle


\noindent{\bf Introduction:} Correlations, often non-local and sometimes surprising, are a salient property of quantum theory. In space-like separated systems, they are quantified by the Bell or Clauser-Horn-Shimony-Holt (CHSH) parameter $B_2$ that encodes joint probabilities of qubit projections onto two distinct sets of non-commuting observables~\cite{chsh1969,Bell2004}. Its excesses above two have been routinely used to test local realism theories~\cite{Hensen2015,2018}. Deeply unsettling to their original investigators~\cite{epr1935}, these nonlocal correlations~\cite{Brunner2014} are now recognized as an important resource for quantum communication, computing, and information tasks~\cite{Barrett2005} including improved odds in nonlocal games~\cite{Palazuelos2016}. For two qubits, the CHSH parameter is restricted to $B_2\leq 2\sqrt{2}$ by the Hilbert-space structure of the quantum theory (Cirel'son bound)~\cite{Cirelson1980,Tsirelson1987}. Quantum measure theory~\cite{Sorkin1994} hypothesizes stronger correlations, for example via Popescu-Rohrlich boxes~\cite{Popescu2014}, but none have been observed. By construction, these single-time correlations are indifferent to the system dynamics. 

Quantum theory also constrains temporal correlations among joint probabilities of a single observable measured at different times~\cite{Leggett1985,Emary2013}. In the minimal model with a single, projectively measured qubit, the correlations are quantified by the Leggett-Garg (LG)  parameter $K_3\equiv C_{21}+C_{32}-C_{31}$, where each $|C_{ij}|\leq 1$, heuristically, encodes the degree of parallelness of the qubit states at times $t_i\geq t_j$. Mathematically, the two-time correlation function is given by $C_{ij}=\sum_{ab}ab P_{ij}(a,b)$ where $a,b=\pm 1$ are two possible outcomes for a dichotomous observable $Q$, and $P_{ij}(a,b)$ is the joint probability of obtaining outcome $b$ at time $t_j$, and after evolving the resulting state from $t_j$ to $t_i$, obtaining an outcome $a$ at time $t_i\geq t_j$. For a two-level system, the maximum possible value of $K_3$ is $K_3^{L}=1.5$, called the L\"{u}der bound~\cite{Budroni2013}. Systems with $K_3>1$, too, have been used to probe local realism theories and associated loopholes~\cite{Knee2016,Joarder2022}. In a multi-level system, these temporal correlations are still bounded by $K_3^L=1.5$~\cite{Budroni2013} for the L\"{u}der (coherent update) protocol~\cite{Luders2006}, whereas with von-Neumann update, $\max K_3\leq 3$ increases with the number of levels~\cite{Budroni2014a,Zhan2023}. 


\begin{figure*}
\centering
\includegraphics[width=\textwidth]{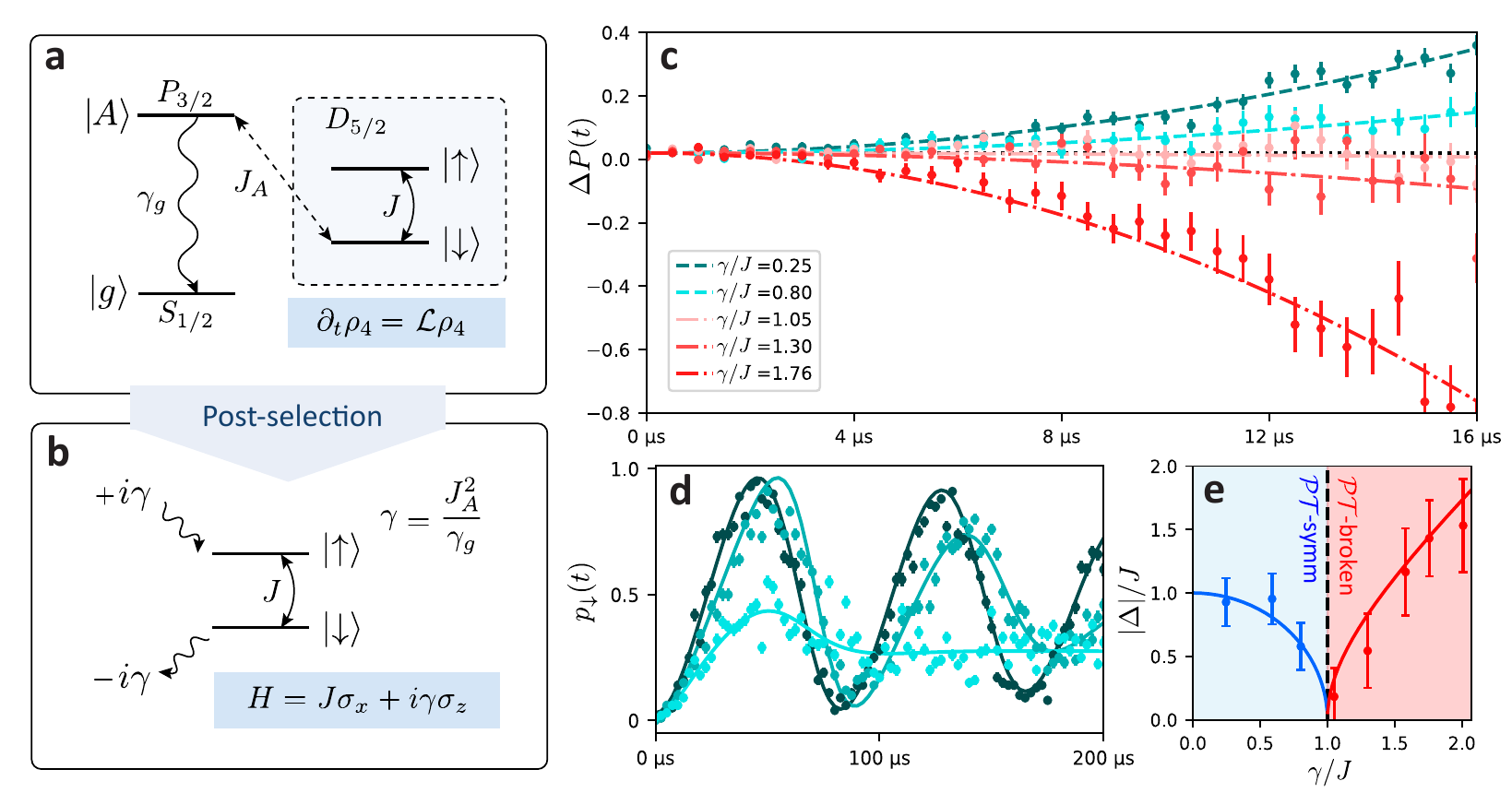}
\caption{\textbf{Two-level non-Hermitian trapped ion}. {\bf a.} Four levels of a $^{40}$Ca$^{+}$ ion $\{\ket{\uparrow},\ket{\downarrow},\ket{A},\ket{g}\}$ have a density matrix $\rho_4(t)$ that is governed by Lindblad equation with two coherent drives and a spontaneous-emission dissipator to the $\ket{g}$ level. {\bf b.} Eliminating the short-lived $\ket{A}$ level and post-selection leads to a non-Hermitian qubit within the $\{\ket{\uparrow},\ket{\downarrow}\}$ manifold; both Rabi drive $J$ and the anti-Hermitian term $\gamma$ in Eq.(\ref{eq:hdimer}) can be independently varied. {\bf c.} Population-transfer difference $\Delta P(t)$ measured at short times $t\lesssim 16$ $\mu$s shows the transition from the $\mathcal{PT}$-symmetric phase ($\Delta P>0$) to a $\mathcal{PT}$-broken phase ($\Delta P<0$), with the EP at $\gamma_\textrm{EP}=J$. {\bf d.} The $\ket{\downarrow}$-level probability $p_{\downarrow}(t)$ measured over 200 $\mu$s shows three skewed oscillations that are increasingly affected by the backflow to the qubit manifold as $\gamma$ increases ($\gamma/J=0.18$ dark green, $\gamma/J=0.37$ medium, $\gamma/J=0.73$ light blue-green) {\bf e.} $|\Delta|$ extracted from parabolic fits to $\Delta P(t)$ shows the expected transition at the EP. (Data: symbols, theory: solid lines, best-fit: broken lines; error bars in {\bf c-d} are s.d. from 400 shots.)} 
\label{fig:one}
\end{figure*}


Temporal correlations, unlike spatial ones, do depend on the dynamics used to evolve the system between two measurements. For a two-level, unitary system $K_3\leq1.5$ is inviolable. Its dissipative interaction with an environment, modeled with Lindblad equation or quantum maps, suppresses $\max K_3$ to smaller values, often into the classical regime $K_3\leq 1$. Here, we show that a two-level system governed by a parity-time ($\mathcal{PT}$)-symmetric Hamiltonian possesses super-quantum correlations, i.e. $K_3>1.5$, across a wide range of non-Hermiticity. This feat is achieved by non-reciprocal transition to an antipodal state in time shorter than $\tau_{\textrm{QSL}}$ decreed by the quantum-speed-limit for a two-level system~\cite{Margolus1998,Levitin2009,Campaioli2019,Ness2021}.    

$\mathcal{PT}$-symmetric Hamiltonians describe open, classical systems with balanced gain and loss~\cite{Bender1998,Joglekar2013,Ashida2020}. Quantum noise in linear amplifiers makes their quantum realization challenging~\cite{Caves1982}. We bypass this barrier by post-selecting on coherent trajectories with no quantum jumps~\cite{Naghiloo2019}. 


\noindent{\bf Non-Hermitian trapped-ion qubit:} We use a single $^{40}$Ca$^{+}$ ion in a linear-Paul trap, with states $\ket{\uparrow} \equiv \ket{m=+ 5/2}$ and $\ket{\downarrow} \equiv\ket{m=+3/2}$ within the meta-stable $D_{5/2}$ manifold as the two-level system (Fig.~\ref{fig:one}a)~\cite{Sherman2013}.  Through post-selection, we realize the non-Hermitian Hamiltonian ($\hbar=1$) 
\begin{align}
    \label{eq:hdimer}
    H(\gamma)=J\sigma_x+i\gamma\sigma_z
\end{align}
as follows (see Methods A, B). The Hermitian Rabi drive $J\sigma_x=J(\ket{\uparrow}\bra{\downarrow}+\ket{\downarrow}\bra{\uparrow})$ is implemented by using resonant radio frequency pulses at the qubit frequency. Additionally, the $\ket{\downarrow}$ state is coupled to the auxiliary, short-lived $P_{3/2}$ state $\ket{A}$ using $\pi$-polarized light with pulse-strength $J_{A}$, where population then primarily (93.5\%) decays to the $S_{1/2}$ ground-state $\ket{g}$ with decay rate $\gamma_{g}$. The dissipative dynamics of the four levels $\{\ket{\uparrow},\ket{\downarrow},\ket{A},\ket{g}\}$ are described by a Lindblad equation with two Hermitian drives $J\sigma_x$ and $J_A(\ket{\uparrow}\bra{A}+\ket{A}\bra{\uparrow})$, and a spontaneous emission dissipator $\sqrt{\gamma_g}\ket{g}\bra{A}$. When $\gamma_g\gg J_{A}$, the auxiliary level can be eliminated and post-selection generates the anti-Hermitian potential $i\gamma\sigma_z=i\gamma(\ket{\uparrow}\bra{\uparrow}-\ket{\downarrow}\bra{\downarrow})$ with $\gamma=J_A^2/\gamma_g\ll\gamma_g$ (Fig.~\ref{fig:one}b). 


\begin{figure*}
\centering
\includegraphics[width=\textwidth]{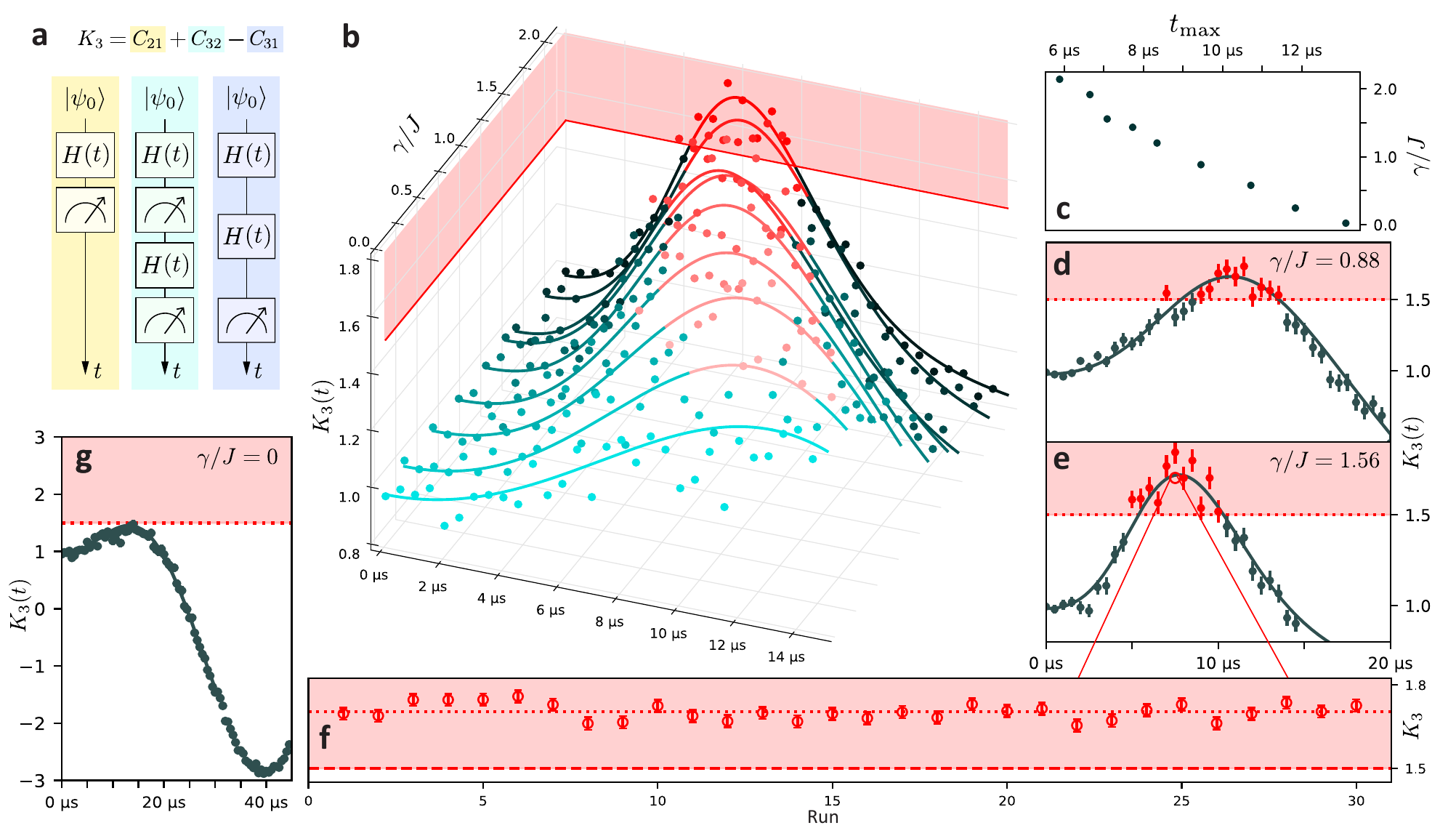}
\caption{\textbf{Super-quantum correlations in $K_3(t)$}. {\bf a.} General protocol for determining $K_3(t)$ comprises three, pairwise two-time projective measurements of a dichotomous observable. {\bf b.} Measured $K_3(t,\gamma)$ exceeds the L\"{u}der bound of 1.5 over a wide range of $\gamma/J$. Different shades of green ($K_3<1.5$) and red ($K_3>1.5$) represent different $\gamma/J$ values. {\bf c.} The $\max K_3$ occurs at time $t_{\max}(\gamma)$ shorter than the unitary-limit value $Jt=\pi/6$. Thus, coherent, non-unitary dynamics generate stronger correlations faster. {\bf d., e.} Time-series for $K_3(t)$ below the EP ($\gamma=0.88J$) and above the EP ($\gamma=1.56J$) show that the location of $\max K_3$ shifts to lower time-values as $\gamma$ is increased. {\bf f.} For a given $(t,\gamma/J)$, the measured $K_3$ remains stable over time across experiments carried out over five hours, thereby showing the robustness of super-quantum correlations in coherent, non-unitary dynamics. A red dotted line shows the average $K_3=1.703(4)$ across these runs. For no run was the L\"{u}der bound of 1.5 exceeded by less than 6.8 standard deviations. {\bf g.} $K_3(t)$ in the unitary case spans the range from 1.5 to -3~\cite{Emary2013}. (data: symbols; theory: lines; each point denotes the average of 400 shots.)}
\label{fig:two}
\end{figure*}

The Hamiltonian $H(\gamma)$ is $\mathcal{PT}$-symmetric with $\mathcal{P}=\sigma_x$ and complex conjugation as the $\mathcal{T}$-operator~\cite{Naghiloo2019}. Its eigenvalues $\pm\Delta=\pm\sqrt{J^2-\gamma^2}$ change from real ($\mathcal{PT}$-symmetric phase) to imaginary ($\mathcal{PT}$-broken phase) at the exceptional-point (EP) degeneracy $\gamma=\gamma_\mathrm{EP}=J$; both $J,\gamma$ are experimentally determined and can be independently controlled. Post-selection preserves the state-norm in the qubit manifold, and leads to a nonlinear equation for the qubit~\cite{Brody2012,Varma2022}
with solution given by 
\begin{align}
\label{eq:gf1}
\ket{\psi(t)}=\frac{G(t)\ket{\psi(0)}}{\sqrt{\bra{\psi(0)}G^\dagger(t)G(t)\ket{\psi(0)}}},
\end{align}
where $G(t)=\cos(\Delta t)\mathbbm{1}_2-iH\sin(\Delta t)/\Delta$ is the non-unitary time-evolution operator. Traditionally, the qubit level probabilities $p_{\downarrow}(t)$ and $p_{\uparrow}(t)=1-p_{\downarrow}(t)$ are measured at times $0\leq t\lesssim T_\Delta=2\pi/|\Delta|$ to characterize the transition across the EP. In our setup, a small fraction (5.87\%) of the population from the auxiliary state decays back into the $D_{5/2}$ manifold (see Methods A). Therefore, at times $t\sim T_{\Delta}$ this backflow creates deviations from Eq.(\ref{eq:hdimer}) as the effective description and makes this approach ill-suited to observe the transition (Fig.~\ref{fig:one}d).  

Instead, we use the sign of the population-transfer difference $\Delta P(t)\equiv P_{\gamma}(t)-P_{J}(t)$ to identify $\mathcal{PT}$-symmetric phase ($\Delta P>0$) and the $\mathcal{PT}$-broken phase ($\Delta P<0$)~\cite{Ding2021}. Here $P_\gamma(t)\equiv|\bra{\downarrow}G(t)\ket{\uparrow}|^2=(J^2/\Delta^2)\sin^2(\Delta t)$ and $P_J(t)\equiv|\bra{-}G(t)\ket{+}|^2=(\gamma^2/\Delta^2)\sin^2(\Delta t)$ denote population transfers to respective antipodal states in the two bases, and $\ket{\pm}\equiv (\ket{\uparrow}\pm\ket{\downarrow})/\sqrt{2}$. Thus $\Delta P(t)=\sin^2(\Delta t)$, obtained at even short times $|\Delta| t\ll 1$, changes sign at the EP and allows determination of $\mathcal{PT}$-symmetry breaking transition. The population transfers $P_\gamma(t),P_J(t)$ are related to the level-transition probabilities $p_{\downarrow},p_{-}$ and the exponentially decaying successful-post-selection fractions $F_{\uparrow},F_{+}$ -- four experimentally measured quantities -- as follows:
\begin{align}
P_\gamma(t)=e^{2\gamma t}F_{\uparrow}(t)p_{\downarrow}(t); \, P_J(t)=e^{2\gamma t}F_{+}(t)p_{-}(t).
\end{align}
Figure ~\ref{fig:one}c shows that the measured $\Delta P(t)$ changes from positive to negative as $\gamma$ traverses across the EP at $\gamma_\textrm{EP}=J$. Detecting this transition through a single-time-instance data in a non-Hermitian qubit is an embodiment of quantum advantage over its classical counterparts that require data over time $t\sim T_{\Delta}$.


\noindent{\bf Super-quantum Leggett-Garg correlations:} The general protocol for measuring the LG parameter $K_3$ is schematically shown in Fig.~\ref{fig:two}a. We use $Q=\sigma_z$ as the dichotomous observable with eigenvalues $\pm 1$ and corresponding one-dimensional projectors $\ket{\uparrow}\bra{\uparrow}$ and $\ket{\downarrow}\bra{\downarrow}$. With initial state $\ket{\psi(0)}=\ket{\downarrow}$ and equally-spaced times $t_1=0$, $t_2=t$, $t_3=2t$, the LG parameter becomes 
\begin{align}
K_3(t,\gamma)&=C(t)+F(t)-C(2t),
\label{eq:lg}
\end{align}
where, unlike the unitary case, $C_{21}=C(t)$ and $C_{32}=F$ are distinct functions  (see Methods C). Starting from one, $K_3(t)$ reaches the L\"{u}der bound of 1.5 at $Jt=\pi/6\approx 0.523$ for a Hermitian qubit. At small $\gamma$, expansion around this point gives $K_3\approx K_3^L+7\sqrt{3}\gamma/(8J)$ thereby exceeding the L\"{u}der bound. Near the EP, similar analysis predicts that $\max K_3$ occurs at $Jt\approx 0.35$ and approaches its algebraic maximum of three in a vanishingly small window at short times $t\propto\gamma^{-1}\ln(2\gamma/J)$ deep in the $\mathcal{PT}$-broken region $\gamma/J\gg 1$. Optimizing $K_3$ over the space of $\{\ket{\psi(0)},Q\}$ significantly broadens this window; the maximum occurs at times longer than the unitary-limit value~\cite{Varma2022} (see Methods C). 


\begin{figure*}
\centering
\includegraphics[width=\textwidth]{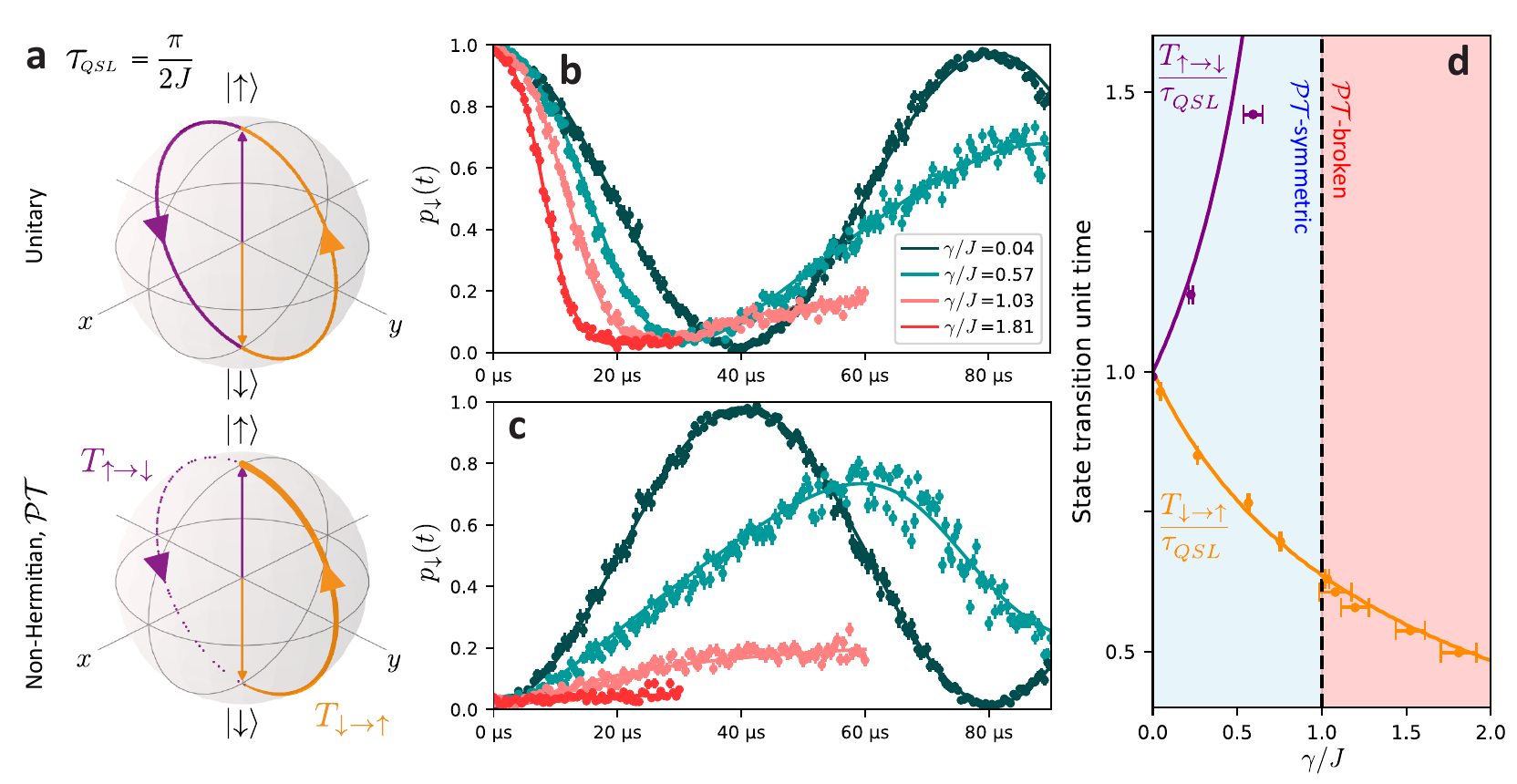}
\caption{{\bf Moving faster than quantum-speed-limit on the Bloch sphere.} {\bf a.} In unitary case, the transit times to and from an antipodal state are equal, bounded below by unified quantum speed limit $\tau_{\text{QSL}}=\pi/(2J)$; non-Hermiticity breaks this symmetry. {\bf b.} Measured $p_\downarrow(t)$ shows a faster transit to the $\ket{\uparrow}$ state as $\gamma$ is increased. {\bf c.} In contrast, $p_{\downarrow}(t)$ shows a slower transit from the $\ket{\uparrow}$ state. The incomplete transfer to the $\ket{\downarrow}$ state also reflects the backflow to the qubit manifold. {\bf d.} Measured transit time $T_{\downarrow\rightarrow\uparrow}$ shows a 50\% reduction relative to its minimum mandated value $\tau_{\text{QSL}}$; the error in measured $T_{\uparrow\rightarrow\downarrow}(\gamma)$ is due to the backflow and the incomplete Rabi flop. (Data: symbols; theory: lines.)}
\label{fig:three}
\end{figure*}


Figure~\ref{fig:two}b shows the experimentally measured $K_3(t,\gamma)$ time-series with $\gamma/J$ values ranging across the EP with $K_3>1.5$ marked red. As $\gamma$ increases, the time at which $K_3$ is maximum shortens, to less than half its unitary-limit value at $\gamma/J\approx 2$ (Fig.~\ref{fig:two}c). A close-up time-series across the EP shows this trend clearly (Fig.~\ref{fig:two}d,e). The coherent, non-unitary dynamics of a $\mathcal{PT}$-symmetric qubit, thus, spawn temporal correlations that are stronger than the L\"{u}der bound, in a time that is shorter that one dictated by the unified quantum speed limit~\cite{Margolus1998,Levitin2009}. 

Accessing deep $\mathcal{PT}$-broken region where $K_3\rightarrow 3$~\cite{Varma2022} is challenging for two reasons. One, increasing $\gamma$ by ramping up the $J_A$ drive suppresses the successful post-selection fraction; two, at a fixed value of $\gamma$, reducing $J$ slow down the dynamics and therefore enhances the effects of backflow. Thus, our observed values reach $\max K_3=1.703(4)$, clearly exceeding the L\"{u}der bound of 1.5.


\noindent{\bf Moving, fast and slow, on the Bloch sphere:} The excesses above 1.5 in $K_3(t)$ arise primarily from the temporal variation of the norm of $G(t)$. For a unitary qubit, the L\"{u}der bound emerges from a constant speed-of-evolution on the Bloch sphere~\cite{Margolus1998,Levitin2009,Campaioli2019}. Being equal for antipodal states, for a two-level system, it results in the unified lower bound $\tau_\textrm{QSL}=\pi/2J$ for the minimum transit-time required to reach an antipodal state (Fig.~\ref{fig:three}a). The coherent, non-unitary dynamics generated by $G(t)$ break this symmetry, and the new transit times satisfy $T_{\downarrow\rightarrow\uparrow}(\gamma)\leq\tau_\textrm{QSL}\leq T_{\uparrow\rightarrow\downarrow}(\gamma)$. As $\gamma$ is increased, $T_{\downarrow\rightarrow\uparrow}$ is suppressed from $\tau_{\textrm{QSL}}$ to $1/J$ at the EP, to $\gamma^{-1}\ln(4\gamma/J)$ for $\gamma\gg J$. The return time satisfies 
\begin{align}
    \label{eq:tupdown}
    T_{\uparrow\rightarrow\downarrow}(\gamma)+T_{\downarrow\rightarrow\uparrow}(\gamma)=\frac{\pi}{\Delta(\gamma)}
\end{align}
when $\gamma< J$, and diverges at and past the EP. 

Measured level occupations $p_{\downarrow}(t)$ show that the Rabi-flop time is shortened for the $\ket{\downarrow}$ state (Fig.~\ref{fig:three}b) and increased for the $\ket{\uparrow}$ state (Fig.~\ref{fig:three}c) as $\gamma/J$ increases. The extracted transit-times $T_{\downarrow\rightarrow\uparrow}(\gamma)$ match theory, showing a 50\% reduction across the EP (Fig.~\ref{fig:three}d); in contrast, transit to the $\ket{\downarrow}$ state from an $\ket{\uparrow}$ state is not observed at large $\gamma$ (Fig.~\ref{fig:three}d). 


\noindent{\bf Discussion:} $\mathcal{PT}$-symmetric Hamiltonians with EP degeneracies~\cite{Miri2019} have shown numerous unusual phenomena such as enhanced sensing~\cite{Hodaei2017,Chen2017}, topological mode switch~\cite{Doppler2016,Xu2016} and non-Hermitian braiding~\cite{Wang2021,Patil2022} in classical platforms. Some of them have been reproduced in quantum platforms via Lindblad post-selection~\cite{Abbasi2022} or unitary dilation~\cite{Wu2019,Liu2021,Maraviglia2022}. However, the interplay between non-Hermiticity and quantum correlations has remained an open question.

By implementing a two-level non-Hermitian Hamiltonian in a single, dissipative ion, we have demonstrated that coherent, non-unitary dynamics create stronger temporal correlations that surpass all known models comprising unitary or dissipative, linear quantum maps~\cite{Budroni2013,Budroni2014a}. When optimized over initial states, these super-quantum correlations are generated faster than those permitted by standard quantum speed limits~\cite{Margolus1998,Levitin2009}. 

These dual features - stronger correlations, faster - of quantum, $\mathcal{PT}$-symmetric Hamiltonians will pave the way for new models of quantum correlations and entanglement generation. 


\begin{acknowledgments}
We thank D.J. Wineland for comments on the manuscript. S.D. thanks Anant Varma for discussions. J.E.M. and Y.N.J. are supported by ONR Grant No. N00014-21-1-2630. A.Q., J.M., I.D.M., S.B., and D.T.C.A. wish to acknowledge support from NSF through the Q-SEnSE Quantum Leap Challenge Institute, Award \#2016244, and the US Army Research Office under award W911NF-20-1-0037.  The data supporting the figures in this article are available upon reasonable request from D.T.C.A. 
\end{acknowledgments}


\bibliography{ptyj1,ptyj2}

\begin{thebibliography}{50}%
\makeatletter
\providecommand \@ifxundefined [1]{%
 \@ifx{#1\undefined}
}%
\providecommand \@ifnum [1]{%
 \ifnum #1\expandafter \@firstoftwo
 \else \expandafter \@secondoftwo
 \fi
}%
\providecommand \@ifx [1]{%
 \ifx #1\expandafter \@firstoftwo
 \else \expandafter \@secondoftwo
 \fi
}%
\providecommand \natexlab [1]{#1}%
\providecommand \enquote  [1]{``#1''}%
\providecommand \bibnamefont  [1]{#1}%
\providecommand \bibfnamefont [1]{#1}%
\providecommand \citenamefont [1]{#1}%
\providecommand \href@noop [0]{\@secondoftwo}%
\providecommand \href [0]{\begingroup \@sanitize@url \@href}%
\providecommand \@href[1]{\@@startlink{#1}\@@href}%
\providecommand \@@href[1]{\endgroup#1\@@endlink}%
\providecommand \@sanitize@url [0]{\catcode `\\12\catcode `\$12\catcode
  `\&12\catcode `\#12\catcode `\^12\catcode `\_12\catcode `\%12\relax}%
\providecommand \@@startlink[1]{}%
\providecommand \@@endlink[0]{}%
\providecommand \url  [0]{\begingroup\@sanitize@url \@url }%
\providecommand \@url [1]{\endgroup\@href {#1}{\urlprefix }}%
\providecommand \urlprefix  [0]{URL }%
\providecommand \Eprint [0]{\href }%
\providecommand \doibase [0]{https://doi.org/}%
\providecommand \selectlanguage [0]{\@gobble}%
\providecommand \bibinfo  [0]{\@secondoftwo}%
\providecommand \bibfield  [0]{\@secondoftwo}%
\providecommand \translation [1]{[#1]}%
\providecommand \BibitemOpen [0]{}%
\providecommand \bibitemStop [0]{}%
\providecommand \bibitemNoStop [0]{.\EOS\space}%
\providecommand \EOS [0]{\spacefactor3000\relax}%
\providecommand \BibitemShut  [1]{\csname bibitem#1\endcsname}%
\let\auto@bib@innerbib\@empty
\bibitem [{\citenamefont {Bell}\ and\ \citenamefont {Aspect}(2004)}]{Bell2004}%
  \BibitemOpen
  \bibfield  {author} {\bibinfo {author} {\bibfnamefont {J.~S.}\ \bibnamefont
  {Bell}}\ and\ \bibinfo {author} {\bibfnamefont {A.}~\bibnamefont {Aspect}},\
  }\href {https://doi.org/10.1017/cbo9780511815676} {\emph {\bibinfo {title}
  {Speakable and Unspeakable in Quantum Mechanics}}}\ (\bibinfo  {publisher}
  {Cambridge University Press},\ \bibinfo {year} {2004})\BibitemShut {NoStop}%
\bibitem [{\citenamefont {Clauser}\ \emph {et~al.}(1969)\citenamefont
  {Clauser}, \citenamefont {Horne}, \citenamefont {Shimony},\ and\
  \citenamefont {Holt}}]{chsh1969}%
  \BibitemOpen
  \bibfield  {author} {\bibinfo {author} {\bibfnamefont {J.~F.}\ \bibnamefont
  {Clauser}}, \bibinfo {author} {\bibfnamefont {M.~A.}\ \bibnamefont {Horne}},
  \bibinfo {author} {\bibfnamefont {A.}~\bibnamefont {Shimony}},\ and\ \bibinfo
  {author} {\bibfnamefont {R.~A.}\ \bibnamefont {Holt}},\ }\bibfield  {title}
  {\bibinfo {title} {Proposed experiment to test local hidden-variable
  theories},\ }\href {https://doi.org/10.1103/PhysRevLett.23.880} {\bibfield
  {journal} {\bibinfo  {journal} {Phys. Rev. Lett.}\ }\textbf {\bibinfo
  {volume} {23}},\ \bibinfo {pages} {880} (\bibinfo {year} {1969})}\BibitemShut
  {NoStop}%
\bibitem [{\citenamefont {Leggett}\ and\ \citenamefont
  {Garg}(1985)}]{Leggett1985}%
  \BibitemOpen
  \bibfield  {author} {\bibinfo {author} {\bibfnamefont {A.~J.}\ \bibnamefont
  {Leggett}}\ and\ \bibinfo {author} {\bibfnamefont {A.}~\bibnamefont {Garg}},\
  }\bibfield  {title} {\bibinfo {title} {Quantum mechanics versus macroscopic
  realism: Is the flux there when nobody looks?},\ }\href
  {https://doi.org/10.1103/PhysRevLett.54.857} {\bibfield  {journal} {\bibinfo
  {journal} {Phys. Rev. Lett.}\ }\textbf {\bibinfo {volume} {54}},\ \bibinfo
  {pages} {857} (\bibinfo {year} {1985})}\BibitemShut {NoStop}%
\bibitem [{\citenamefont {Emary}\ \emph {et~al.}(2013)\citenamefont {Emary},
  \citenamefont {Lambert},\ and\ \citenamefont {Nori}}]{Emary2013}%
  \BibitemOpen
  \bibfield  {author} {\bibinfo {author} {\bibfnamefont {C.}~\bibnamefont
  {Emary}}, \bibinfo {author} {\bibfnamefont {N.}~\bibnamefont {Lambert}},\
  and\ \bibinfo {author} {\bibfnamefont {F.}~\bibnamefont {Nori}},\ }\bibfield
  {title} {\bibinfo {title} {Leggett{\textendash}garg inequalities},\ }\href
  {https://doi.org/10.1088/0034-4885/77/1/016001} {\bibfield  {journal}
  {\bibinfo  {journal} {Reports on Progress in Physics}\ }\textbf {\bibinfo
  {volume} {77}},\ \bibinfo {pages} {016001} (\bibinfo {year}
  {2013})}\BibitemShut {NoStop}%
\bibitem [{\citenamefont {Hensen}\ \emph {et~al.}(2015)\citenamefont {Hensen},
  \citenamefont {Bernien}, \citenamefont {Dr{\'{e}}au}, \citenamefont
  {Reiserer}, \citenamefont {Kalb}, \citenamefont {Blok}, \citenamefont
  {Ruitenberg}, \citenamefont {Vermeulen}, \citenamefont {Schouten},
  \citenamefont {Abell{\'{a}}n}, \citenamefont {Amaya}, \citenamefont
  {Pruneri}, \citenamefont {Mitchell}, \citenamefont {Markham}, \citenamefont
  {Twitchen}, \citenamefont {Elkouss}, \citenamefont {Wehner}, \citenamefont
  {Taminiau},\ and\ \citenamefont {Hanson}}]{Hensen2015}%
  \BibitemOpen
  \bibfield  {author} {\bibinfo {author} {\bibfnamefont {B.}~\bibnamefont
  {Hensen}}, \bibinfo {author} {\bibfnamefont {H.}~\bibnamefont {Bernien}},
  \bibinfo {author} {\bibfnamefont {A.~E.}\ \bibnamefont {Dr{\'{e}}au}},
  \bibinfo {author} {\bibfnamefont {A.}~\bibnamefont {Reiserer}}, \bibinfo
  {author} {\bibfnamefont {N.}~\bibnamefont {Kalb}}, \bibinfo {author}
  {\bibfnamefont {M.~S.}\ \bibnamefont {Blok}}, \bibinfo {author}
  {\bibfnamefont {J.}~\bibnamefont {Ruitenberg}}, \bibinfo {author}
  {\bibfnamefont {R.~F.~L.}\ \bibnamefont {Vermeulen}}, \bibinfo {author}
  {\bibfnamefont {R.~N.}\ \bibnamefont {Schouten}}, \bibinfo {author}
  {\bibfnamefont {C.}~\bibnamefont {Abell{\'{a}}n}}, \bibinfo {author}
  {\bibfnamefont {W.}~\bibnamefont {Amaya}}, \bibinfo {author} {\bibfnamefont
  {V.}~\bibnamefont {Pruneri}}, \bibinfo {author} {\bibfnamefont {M.~W.}\
  \bibnamefont {Mitchell}}, \bibinfo {author} {\bibfnamefont {M.}~\bibnamefont
  {Markham}}, \bibinfo {author} {\bibfnamefont {D.~J.}\ \bibnamefont
  {Twitchen}}, \bibinfo {author} {\bibfnamefont {D.}~\bibnamefont {Elkouss}},
  \bibinfo {author} {\bibfnamefont {S.}~\bibnamefont {Wehner}}, \bibinfo
  {author} {\bibfnamefont {T.~H.}\ \bibnamefont {Taminiau}},\ and\ \bibinfo
  {author} {\bibfnamefont {R.}~\bibnamefont {Hanson}},\ }\bibfield  {title}
  {\bibinfo {title} {Loophole-free bell inequality violation using electron
  spins separated by 1.3 kilometres},\ }\href
  {https://doi.org/10.1038/nature15759} {\bibfield  {journal} {\bibinfo
  {journal} {Nature}\ }\textbf {\bibinfo {volume} {526}},\ \bibinfo {pages}
  {682} (\bibinfo {year} {2015})}\BibitemShut {NoStop}%
\bibitem [{201(2018)}]{2018}%
  \BibitemOpen
  \bibfield  {title} {\bibinfo {title} {Challenging local realism with human
  choices},\ }\href {https://doi.org/10.1038/s41586-018-0085-3} {\bibfield
  {journal} {\bibinfo  {journal} {Nature}\ }\textbf {\bibinfo {volume} {557}},\
  \bibinfo {pages} {212} (\bibinfo {year} {2018})}\BibitemShut {NoStop}%
\bibitem [{\citenamefont {Palacios-Laloy}\ \emph {et~al.}(2010)\citenamefont
  {Palacios-Laloy}, \citenamefont {Mallet}, \citenamefont {Nguyen},
  \citenamefont {Bertet}, \citenamefont {Vion}, \citenamefont {Esteve},\ and\
  \citenamefont {Korotkov}}]{PalaciosLaloy2010}%
  \BibitemOpen
  \bibfield  {author} {\bibinfo {author} {\bibfnamefont {A.}~\bibnamefont
  {Palacios-Laloy}}, \bibinfo {author} {\bibfnamefont {F.}~\bibnamefont
  {Mallet}}, \bibinfo {author} {\bibfnamefont {F.}~\bibnamefont {Nguyen}},
  \bibinfo {author} {\bibfnamefont {P.}~\bibnamefont {Bertet}}, \bibinfo
  {author} {\bibfnamefont {D.}~\bibnamefont {Vion}}, \bibinfo {author}
  {\bibfnamefont {D.}~\bibnamefont {Esteve}},\ and\ \bibinfo {author}
  {\bibfnamefont {A.~N.}\ \bibnamefont {Korotkov}},\ }\bibfield  {title}
  {\bibinfo {title} {Experimental violation of a bell's inequality in time with
  weak measurement},\ }\href {https://doi.org/10.1038/nphys1641} {\bibfield
  {journal} {\bibinfo  {journal} {Nature Physics}\ }\textbf {\bibinfo {volume}
  {6}},\ \bibinfo {pages} {442} (\bibinfo {year} {2010})}\BibitemShut {NoStop}%
\bibitem [{\citenamefont {Athalye}\ \emph {et~al.}(2011)\citenamefont
  {Athalye}, \citenamefont {Roy},\ and\ \citenamefont {Mahesh}}]{Athalye2011}%
  \BibitemOpen
  \bibfield  {author} {\bibinfo {author} {\bibfnamefont {V.}~\bibnamefont
  {Athalye}}, \bibinfo {author} {\bibfnamefont {S.~S.}\ \bibnamefont {Roy}},\
  and\ \bibinfo {author} {\bibfnamefont {T.~S.}\ \bibnamefont {Mahesh}},\
  }\bibfield  {title} {\bibinfo {title} {Investigation of the leggett-garg
  inequality for precessing nuclear spins},\ }\href
  {https://doi.org/10.1103/PhysRevLett.107.130402} {\bibfield  {journal}
  {\bibinfo  {journal} {Phys. Rev. Lett.}\ }\textbf {\bibinfo {volume} {107}},\
  \bibinfo {pages} {130402} (\bibinfo {year} {2011})}\BibitemShut {NoStop}%
\bibitem [{\citenamefont {Waldherr}\ \emph {et~al.}(2011)\citenamefont
  {Waldherr}, \citenamefont {Neumann}, \citenamefont {Huelga}, \citenamefont
  {Jelezko},\ and\ \citenamefont {Wrachtrup}}]{Waldherr2011}%
  \BibitemOpen
  \bibfield  {author} {\bibinfo {author} {\bibfnamefont {G.}~\bibnamefont
  {Waldherr}}, \bibinfo {author} {\bibfnamefont {P.}~\bibnamefont {Neumann}},
  \bibinfo {author} {\bibfnamefont {S.~F.}\ \bibnamefont {Huelga}}, \bibinfo
  {author} {\bibfnamefont {F.}~\bibnamefont {Jelezko}},\ and\ \bibinfo {author}
  {\bibfnamefont {J.}~\bibnamefont {Wrachtrup}},\ }\bibfield  {title} {\bibinfo
  {title} {Violation of a temporal bell inequality for single spins in a
  diamond defect center},\ }\href
  {https://doi.org/10.1103/PhysRevLett.107.090401} {\bibfield  {journal}
  {\bibinfo  {journal} {Phys. Rev. Lett.}\ }\textbf {\bibinfo {volume} {107}},\
  \bibinfo {pages} {090401} (\bibinfo {year} {2011})}\BibitemShut {NoStop}%
\bibitem [{\citenamefont {Goggin}\ \emph {et~al.}(2011)\citenamefont {Goggin},
  \citenamefont {Almeida}, \citenamefont {Barbieri}, \citenamefont {Lanyon},
  \citenamefont {O'Brien}, \citenamefont {White},\ and\ \citenamefont
  {Pryde}}]{Goggin2011}%
  \BibitemOpen
  \bibfield  {author} {\bibinfo {author} {\bibfnamefont {M.~E.}\ \bibnamefont
  {Goggin}}, \bibinfo {author} {\bibfnamefont {M.~P.}\ \bibnamefont {Almeida}},
  \bibinfo {author} {\bibfnamefont {M.}~\bibnamefont {Barbieri}}, \bibinfo
  {author} {\bibfnamefont {B.~P.}\ \bibnamefont {Lanyon}}, \bibinfo {author}
  {\bibfnamefont {J.~L.}\ \bibnamefont {O'Brien}}, \bibinfo {author}
  {\bibfnamefont {A.~G.}\ \bibnamefont {White}},\ and\ \bibinfo {author}
  {\bibfnamefont {G.~J.}\ \bibnamefont {Pryde}},\ }\bibfield  {title} {\bibinfo
  {title} {Violation of the leggett{\textendash}garg inequality with weak
  measurements of photons},\ }\href {https://doi.org/10.1073/pnas.1005774108}
  {\bibfield  {journal} {\bibinfo  {journal} {Proceedings of the National
  Academy of Sciences}\ }\textbf {\bibinfo {volume} {108}},\ \bibinfo {pages}
  {1256} (\bibinfo {year} {2011})}\BibitemShut {NoStop}%
\bibitem [{\citenamefont {Knee}\ \emph {et~al.}(2016)\citenamefont {Knee},
  \citenamefont {Kakuyanagi}, \citenamefont {Yeh}, \citenamefont {Matsuzaki},
  \citenamefont {Toida}, \citenamefont {Yamaguchi}, \citenamefont {Saito},
  \citenamefont {Leggett},\ and\ \citenamefont {Munro}}]{Knee2016}%
  \BibitemOpen
  \bibfield  {author} {\bibinfo {author} {\bibfnamefont {G.~C.}\ \bibnamefont
  {Knee}}, \bibinfo {author} {\bibfnamefont {K.}~\bibnamefont {Kakuyanagi}},
  \bibinfo {author} {\bibfnamefont {M.-C.}\ \bibnamefont {Yeh}}, \bibinfo
  {author} {\bibfnamefont {Y.}~\bibnamefont {Matsuzaki}}, \bibinfo {author}
  {\bibfnamefont {H.}~\bibnamefont {Toida}}, \bibinfo {author} {\bibfnamefont
  {H.}~\bibnamefont {Yamaguchi}}, \bibinfo {author} {\bibfnamefont
  {S.}~\bibnamefont {Saito}}, \bibinfo {author} {\bibfnamefont {A.~J.}\
  \bibnamefont {Leggett}},\ and\ \bibinfo {author} {\bibfnamefont {W.~J.}\
  \bibnamefont {Munro}},\ }\bibfield  {title} {\bibinfo {title} {A strict
  experimental test of macroscopic realism in a superconducting flux qubit},\
  }\bibfield  {journal} {\bibinfo  {journal} {Nature Communications}\ }\textbf
  {\bibinfo {volume} {7}},\ \href {https://doi.org/10.1038/ncomms13253}
  {10.1038/ncomms13253} (\bibinfo {year} {2016})\BibitemShut {NoStop}%
\bibitem [{\citenamefont {Einstein}\ \emph {et~al.}(1935)\citenamefont
  {Einstein}, \citenamefont {Podolsky},\ and\ \citenamefont {Rosen}}]{epr1935}%
  \BibitemOpen
  \bibfield  {author} {\bibinfo {author} {\bibfnamefont {A.}~\bibnamefont
  {Einstein}}, \bibinfo {author} {\bibfnamefont {B.}~\bibnamefont {Podolsky}},\
  and\ \bibinfo {author} {\bibfnamefont {N.}~\bibnamefont {Rosen}},\ }\bibfield
   {title} {\bibinfo {title} {Can quantum-mechanical description of physical
  reality be considered complete?},\ }\href
  {https://doi.org/10.1103/PhysRev.47.777} {\bibfield  {journal} {\bibinfo
  {journal} {Phys. Rev.}\ }\textbf {\bibinfo {volume} {47}},\ \bibinfo {pages}
  {777} (\bibinfo {year} {1935})}\BibitemShut {NoStop}%
\bibitem [{\citenamefont {Brunner}\ \emph {et~al.}(2014)\citenamefont
  {Brunner}, \citenamefont {Cavalcanti}, \citenamefont {Pironio}, \citenamefont
  {Scarani},\ and\ \citenamefont {Wehner}}]{Brunner2014}%
  \BibitemOpen
  \bibfield  {author} {\bibinfo {author} {\bibfnamefont {N.}~\bibnamefont
  {Brunner}}, \bibinfo {author} {\bibfnamefont {D.}~\bibnamefont {Cavalcanti}},
  \bibinfo {author} {\bibfnamefont {S.}~\bibnamefont {Pironio}}, \bibinfo
  {author} {\bibfnamefont {V.}~\bibnamefont {Scarani}},\ and\ \bibinfo {author}
  {\bibfnamefont {S.}~\bibnamefont {Wehner}},\ }\bibfield  {title} {\bibinfo
  {title} {Bell nonlocality},\ }\href
  {https://doi.org/10.1103/RevModPhys.86.419} {\bibfield  {journal} {\bibinfo
  {journal} {Rev. Mod. Phys.}\ }\textbf {\bibinfo {volume} {86}},\ \bibinfo
  {pages} {419} (\bibinfo {year} {2014})}\BibitemShut {NoStop}%
\bibitem [{\citenamefont {Barrett}\ \emph {et~al.}(2005)\citenamefont
  {Barrett}, \citenamefont {Linden}, \citenamefont {Massar}, \citenamefont
  {Pironio}, \citenamefont {Popescu},\ and\ \citenamefont
  {Roberts}}]{Barrett2005}%
  \BibitemOpen
  \bibfield  {author} {\bibinfo {author} {\bibfnamefont {J.}~\bibnamefont
  {Barrett}}, \bibinfo {author} {\bibfnamefont {N.}~\bibnamefont {Linden}},
  \bibinfo {author} {\bibfnamefont {S.}~\bibnamefont {Massar}}, \bibinfo
  {author} {\bibfnamefont {S.}~\bibnamefont {Pironio}}, \bibinfo {author}
  {\bibfnamefont {S.}~\bibnamefont {Popescu}},\ and\ \bibinfo {author}
  {\bibfnamefont {D.}~\bibnamefont {Roberts}},\ }\bibfield  {title} {\bibinfo
  {title} {Nonlocal correlations as an information-theoretic resource},\ }\href
  {https://doi.org/10.1103/PhysRevA.71.022101} {\bibfield  {journal} {\bibinfo
  {journal} {Phys. Rev. A}\ }\textbf {\bibinfo {volume} {71}},\ \bibinfo
  {pages} {022101} (\bibinfo {year} {2005})}\BibitemShut {NoStop}%
\bibitem [{\citenamefont {Palazuelos}\ and\ \citenamefont
  {Vidick}(2016)}]{Palazuelos2016}%
  \BibitemOpen
  \bibfield  {author} {\bibinfo {author} {\bibfnamefont {C.}~\bibnamefont
  {Palazuelos}}\ and\ \bibinfo {author} {\bibfnamefont {T.}~\bibnamefont
  {Vidick}},\ }\bibfield  {title} {\bibinfo {title} {Survey on nonlocal games
  and operator space theory},\ }\href {https://doi.org/10.1063/1.4938052}
  {\bibfield  {journal} {\bibinfo  {journal} {Journal of Mathematical Physics}\
  }\textbf {\bibinfo {volume} {57}},\ \bibinfo {pages} {015220} (\bibinfo
  {year} {2016})}\BibitemShut {NoStop}%
\bibitem [{\citenamefont {Cirel{\textquotesingle}son}(1980)}]{Cirelson1980}%
  \BibitemOpen
  \bibfield  {author} {\bibinfo {author} {\bibfnamefont {B.~S.}\ \bibnamefont
  {Cirel{\textquotesingle}son}},\ }\bibfield  {title} {\bibinfo {title}
  {Quantum generalizations of bell{\textquotesingle}s inequality},\ }\href
  {https://doi.org/10.1007/bf00417500} {\bibfield  {journal} {\bibinfo
  {journal} {Letters in Mathematical Physics}\ }\textbf {\bibinfo {volume}
  {4}},\ \bibinfo {pages} {93} (\bibinfo {year} {1980})}\BibitemShut {NoStop}%
\bibitem [{\citenamefont {Tsirel{\textquotesingle}son}(1987)}]{Tsirelson1987}%
  \BibitemOpen
  \bibfield  {author} {\bibinfo {author} {\bibfnamefont {B.~S.}\ \bibnamefont
  {Tsirel{\textquotesingle}son}},\ }\bibfield  {title} {\bibinfo {title}
  {Quantum analogues of the bell inequalities. the case of two spatially
  separated domains},\ }\href {https://doi.org/10.1007/bf01663472} {\bibfield
  {journal} {\bibinfo  {journal} {Journal of Soviet Mathematics}\ }\textbf
  {\bibinfo {volume} {36}},\ \bibinfo {pages} {557} (\bibinfo {year}
  {1987})}\BibitemShut {NoStop}%
\bibitem [{\citenamefont {Sorkin}(1994)}]{Sorkin1994}%
  \BibitemOpen
  \bibfield  {author} {\bibinfo {author} {\bibfnamefont {R.~D.}\ \bibnamefont
  {Sorkin}},\ }\bibfield  {title} {\bibinfo {title} {Quantum mechanics as
  quantum measure theory},\ }\href {https://doi.org/10.1142/s021773239400294x}
  {\bibfield  {journal} {\bibinfo  {journal} {Modern Physics Letters A}\
  }\textbf {\bibinfo {volume} {09}},\ \bibinfo {pages} {3119} (\bibinfo {year}
  {1994})}\BibitemShut {NoStop}%
\bibitem [{\citenamefont {Popescu}(2014)}]{Popescu2014}%
  \BibitemOpen
  \bibfield  {author} {\bibinfo {author} {\bibfnamefont {S.}~\bibnamefont
  {Popescu}},\ }\bibfield  {title} {\bibinfo {title} {Nonlocality beyond
  quantum mechanics},\ }\href {https://doi.org/10.1038/nphys2916} {\bibfield
  {journal} {\bibinfo  {journal} {Nature Physics}\ }\textbf {\bibinfo {volume}
  {10}},\ \bibinfo {pages} {264} (\bibinfo {year} {2014})}\BibitemShut
  {NoStop}%
\bibitem [{\citenamefont {Budroni}\ \emph {et~al.}(2013)\citenamefont
  {Budroni}, \citenamefont {Moroder}, \citenamefont {Kleinmann},\ and\
  \citenamefont {G\"uhne}}]{Budroni2013}%
  \BibitemOpen
  \bibfield  {author} {\bibinfo {author} {\bibfnamefont {C.}~\bibnamefont
  {Budroni}}, \bibinfo {author} {\bibfnamefont {T.}~\bibnamefont {Moroder}},
  \bibinfo {author} {\bibfnamefont {M.}~\bibnamefont {Kleinmann}},\ and\
  \bibinfo {author} {\bibfnamefont {O.}~\bibnamefont {G\"uhne}},\ }\bibfield
  {title} {\bibinfo {title} {Bounding temporal quantum correlations},\ }\href
  {https://doi.org/10.1103/PhysRevLett.111.020403} {\bibfield  {journal}
  {\bibinfo  {journal} {Phys. Rev. Lett.}\ }\textbf {\bibinfo {volume} {111}},\
  \bibinfo {pages} {020403} (\bibinfo {year} {2013})}\BibitemShut {NoStop}%
\bibitem [{\citenamefont {Joarder}\ \emph {et~al.}(2022)\citenamefont
  {Joarder}, \citenamefont {Saha}, \citenamefont {Home},\ and\ \citenamefont
  {Sinha}}]{Joarder2022}%
  \BibitemOpen
  \bibfield  {author} {\bibinfo {author} {\bibfnamefont {K.}~\bibnamefont
  {Joarder}}, \bibinfo {author} {\bibfnamefont {D.}~\bibnamefont {Saha}},
  \bibinfo {author} {\bibfnamefont {D.}~\bibnamefont {Home}},\ and\ \bibinfo
  {author} {\bibfnamefont {U.}~\bibnamefont {Sinha}},\ }\bibfield  {title}
  {\bibinfo {title} {Loophole-free interferometric test of macrorealism using
  heralded single photons},\ }\href
  {https://doi.org/10.1103/PRXQuantum.3.010307} {\bibfield  {journal} {\bibinfo
   {journal} {PRX Quantum}\ }\textbf {\bibinfo {volume} {3}},\ \bibinfo {pages}
  {010307} (\bibinfo {year} {2022})}\BibitemShut {NoStop}%
\bibitem [{\citenamefont {L\"{u}ders}(2006)}]{Luders2006}%
  \BibitemOpen
  \bibfield  {author} {\bibinfo {author} {\bibfnamefont {G.}~\bibnamefont
  {L\"{u}ders}},\ }\bibfield  {title} {\bibinfo {title} {Concerning the
  state-change due to the measurement process},\ }\href
  {https://doi.org/10.1002/andp.20065180904} {\bibfield  {journal} {\bibinfo
  {journal} {Annalen der Physik}\ }\textbf {\bibinfo {volume} {518}},\ \bibinfo
  {pages} {663} (\bibinfo {year} {2006})}\BibitemShut {NoStop}%
\bibitem [{\citenamefont {Budroni}\ and\ \citenamefont
  {Emary}(2014)}]{Budroni2014a}%
  \BibitemOpen
  \bibfield  {author} {\bibinfo {author} {\bibfnamefont {C.}~\bibnamefont
  {Budroni}}\ and\ \bibinfo {author} {\bibfnamefont {C.}~\bibnamefont
  {Emary}},\ }\bibfield  {title} {\bibinfo {title} {Temporal quantum
  correlations and leggett-garg inequalities in multilevel systems},\ }\href
  {https://doi.org/10.1103/PhysRevLett.113.050401} {\bibfield  {journal}
  {\bibinfo  {journal} {Phys. Rev. Lett.}\ }\textbf {\bibinfo {volume} {113}},\
  \bibinfo {pages} {050401} (\bibinfo {year} {2014})}\BibitemShut {NoStop}%
\bibitem [{\citenamefont {Zhan}\ \emph {et~al.}(2023)\citenamefont {Zhan},
  \citenamefont {Wu}, \citenamefont {Zhang}, \citenamefont {Qin}, \citenamefont
  {Yang}, \citenamefont {Hu}, \citenamefont {Su}, \citenamefont {Zhang},
  \citenamefont {Chen}, \citenamefont {Xie}, \citenamefont {Wu},\ and\
  \citenamefont {Chen}}]{Zhan2023}%
  \BibitemOpen
  \bibfield  {author} {\bibinfo {author} {\bibfnamefont {T.}~\bibnamefont
  {Zhan}}, \bibinfo {author} {\bibfnamefont {C.}~\bibnamefont {Wu}}, \bibinfo
  {author} {\bibfnamefont {M.}~\bibnamefont {Zhang}}, \bibinfo {author}
  {\bibfnamefont {Q.}~\bibnamefont {Qin}}, \bibinfo {author} {\bibfnamefont
  {X.}~\bibnamefont {Yang}}, \bibinfo {author} {\bibfnamefont {H.}~\bibnamefont
  {Hu}}, \bibinfo {author} {\bibfnamefont {W.}~\bibnamefont {Su}}, \bibinfo
  {author} {\bibfnamefont {J.}~\bibnamefont {Zhang}}, \bibinfo {author}
  {\bibfnamefont {T.}~\bibnamefont {Chen}}, \bibinfo {author} {\bibfnamefont
  {Y.}~\bibnamefont {Xie}}, \bibinfo {author} {\bibfnamefont {W.}~\bibnamefont
  {Wu}},\ and\ \bibinfo {author} {\bibfnamefont {P.}~\bibnamefont {Chen}},\
  }\bibfield  {title} {\bibinfo {title} {Experimental violation of the
  leggett-garg inequality in a three-level trapped-ion system},\ }\href
  {https://doi.org/10.1103/PhysRevA.107.012424} {\bibfield  {journal} {\bibinfo
   {journal} {Phys. Rev. A}\ }\textbf {\bibinfo {volume} {107}},\ \bibinfo
  {pages} {012424} (\bibinfo {year} {2023})}\BibitemShut {NoStop}%
\bibitem [{\citenamefont {Margolus}\ and\ \citenamefont
  {Levitin}(1998)}]{Margolus1998}%
  \BibitemOpen
  \bibfield  {author} {\bibinfo {author} {\bibfnamefont {N.}~\bibnamefont
  {Margolus}}\ and\ \bibinfo {author} {\bibfnamefont {L.~B.}\ \bibnamefont
  {Levitin}},\ }\bibfield  {title} {\bibinfo {title} {The maximum speed of
  dynamical evolution},\ }\href {https://doi.org/10.1016/s0167-2789(98)00054-2}
  {\bibfield  {journal} {\bibinfo  {journal} {Physica D: Nonlinear Phenomena}\
  }\textbf {\bibinfo {volume} {120}},\ \bibinfo {pages} {188} (\bibinfo {year}
  {1998})}\BibitemShut {NoStop}%
\bibitem [{\citenamefont {Levitin}\ and\ \citenamefont
  {Toffoli}(2009)}]{Levitin2009}%
  \BibitemOpen
  \bibfield  {author} {\bibinfo {author} {\bibfnamefont {L.~B.}\ \bibnamefont
  {Levitin}}\ and\ \bibinfo {author} {\bibfnamefont {T.}~\bibnamefont
  {Toffoli}},\ }\bibfield  {title} {\bibinfo {title} {Fundamental limit on the
  rate of quantum dynamics: The unified bound is tight},\ }\href
  {https://doi.org/10.1103/PhysRevLett.103.160502} {\bibfield  {journal}
  {\bibinfo  {journal} {Phys. Rev. Lett.}\ }\textbf {\bibinfo {volume} {103}},\
  \bibinfo {pages} {160502} (\bibinfo {year} {2009})}\BibitemShut {NoStop}%
\bibitem [{\citenamefont {Campaioli}\ \emph {et~al.}(2019)\citenamefont
  {Campaioli}, \citenamefont {Pollock},\ and\ \citenamefont
  {Modi}}]{Campaioli2019}%
  \BibitemOpen
  \bibfield  {author} {\bibinfo {author} {\bibfnamefont {F.}~\bibnamefont
  {Campaioli}}, \bibinfo {author} {\bibfnamefont {F.~A.}\ \bibnamefont
  {Pollock}},\ and\ \bibinfo {author} {\bibfnamefont {K.}~\bibnamefont
  {Modi}},\ }\bibfield  {title} {\bibinfo {title} {Tight, robust, and feasible
  quantum speed limits for open dynamics},\ }\href
  {https://doi.org/10.22331/q-2019-08-05-168} {\bibfield  {journal} {\bibinfo
  {journal} {Quantum}\ }\textbf {\bibinfo {volume} {3}},\ \bibinfo {pages}
  {168} (\bibinfo {year} {2019})}\BibitemShut {NoStop}%
\bibitem [{\citenamefont {Ness}\ \emph {et~al.}(2021)\citenamefont {Ness},
  \citenamefont {Lam}, \citenamefont {Alt}, \citenamefont {Meschede},
  \citenamefont {Sagi},\ and\ \citenamefont {Alberti}}]{Ness2021}%
  \BibitemOpen
  \bibfield  {author} {\bibinfo {author} {\bibfnamefont {G.}~\bibnamefont
  {Ness}}, \bibinfo {author} {\bibfnamefont {M.~R.}\ \bibnamefont {Lam}},
  \bibinfo {author} {\bibfnamefont {W.}~\bibnamefont {Alt}}, \bibinfo {author}
  {\bibfnamefont {D.}~\bibnamefont {Meschede}}, \bibinfo {author}
  {\bibfnamefont {Y.}~\bibnamefont {Sagi}},\ and\ \bibinfo {author}
  {\bibfnamefont {A.}~\bibnamefont {Alberti}},\ }\bibfield  {title} {\bibinfo
  {title} {Observing crossover between quantum speed limits},\ }\bibfield
  {journal} {\bibinfo  {journal} {Science Advances}\ }\textbf {\bibinfo
  {volume} {7}},\ \href {https://doi.org/10.1126/sciadv.abj9119}
  {10.1126/sciadv.abj9119} (\bibinfo {year} {2021})\BibitemShut {NoStop}%
\bibitem [{\citenamefont {Bender}\ and\ \citenamefont
  {Boettcher}(1998)}]{Bender1998}%
  \BibitemOpen
  \bibfield  {author} {\bibinfo {author} {\bibfnamefont {C.~M.}\ \bibnamefont
  {Bender}}\ and\ \bibinfo {author} {\bibfnamefont {S.}~\bibnamefont
  {Boettcher}},\ }\bibfield  {title} {\bibinfo {title} {Real spectra in
  non-hermitian hamiltonians having $\mathcal{PT}$ symmetry},\ }\href
  {https://doi.org/10.1103/PhysRevLett.80.5243} {\bibfield  {journal} {\bibinfo
   {journal} {Phys. Rev. Lett.}\ }\textbf {\bibinfo {volume} {80}},\ \bibinfo
  {pages} {5243} (\bibinfo {year} {1998})}\BibitemShut {NoStop}%
\bibitem [{\citenamefont {Joglekar}\ \emph {et~al.}(2013)\citenamefont
  {Joglekar}, \citenamefont {Thompson}, \citenamefont {Scott},\ and\
  \citenamefont {Vemuri}}]{Joglekar2013}%
  \BibitemOpen
  \bibfield  {author} {\bibinfo {author} {\bibfnamefont {Y.~N.}\ \bibnamefont
  {Joglekar}}, \bibinfo {author} {\bibfnamefont {C.}~\bibnamefont {Thompson}},
  \bibinfo {author} {\bibfnamefont {D.~D.}\ \bibnamefont {Scott}},\ and\
  \bibinfo {author} {\bibfnamefont {G.}~\bibnamefont {Vemuri}},\ }\bibfield
  {title} {\bibinfo {title} {Optical waveguide arrays: quantum effects and {PT}
  symmetry breaking},\ }\href {https://doi.org/10.1051/epjap/2013130240}
  {\bibfield  {journal} {\bibinfo  {journal} {The European Physical Journal
  Applied Physics}\ }\textbf {\bibinfo {volume} {63}},\ \bibinfo {pages}
  {30001} (\bibinfo {year} {2013})}\BibitemShut {NoStop}%
\bibitem [{\citenamefont {Ashida}\ \emph {et~al.}(2020)\citenamefont {Ashida},
  \citenamefont {Gong},\ and\ \citenamefont {Ueda}}]{Ashida2020}%
  \BibitemOpen
  \bibfield  {author} {\bibinfo {author} {\bibfnamefont {Y.}~\bibnamefont
  {Ashida}}, \bibinfo {author} {\bibfnamefont {Z.}~\bibnamefont {Gong}},\ and\
  \bibinfo {author} {\bibfnamefont {M.}~\bibnamefont {Ueda}},\ }\bibfield
  {title} {\bibinfo {title} {Non-hermitian physics},\ }\href
  {https://doi.org/10.1080/00018732.2021.1876991} {\bibfield  {journal}
  {\bibinfo  {journal} {Advances in Physics}\ }\textbf {\bibinfo {volume}
  {69}},\ \bibinfo {pages} {249} (\bibinfo {year} {2020})}\BibitemShut
  {NoStop}%
\bibitem [{\citenamefont {Caves}(1982)}]{Caves1982}%
  \BibitemOpen
  \bibfield  {author} {\bibinfo {author} {\bibfnamefont {C.~M.}\ \bibnamefont
  {Caves}},\ }\bibfield  {title} {\bibinfo {title} {Quantum limits on noise in
  linear amplifiers},\ }\href {https://doi.org/10.1103/physrevd.26.1817}
  {\bibfield  {journal} {\bibinfo  {journal} {Physical Review D}\ }\textbf
  {\bibinfo {volume} {26}},\ \bibinfo {pages} {1817} (\bibinfo {year}
  {1982})}\BibitemShut {NoStop}%
\bibitem [{\citenamefont {Naghiloo}\ \emph {et~al.}(2019)\citenamefont
  {Naghiloo}, \citenamefont {Abbasi}, \citenamefont {Joglekar},\ and\
  \citenamefont {Murch}}]{Naghiloo2019}%
  \BibitemOpen
  \bibfield  {author} {\bibinfo {author} {\bibfnamefont {M.}~\bibnamefont
  {Naghiloo}}, \bibinfo {author} {\bibfnamefont {M.}~\bibnamefont {Abbasi}},
  \bibinfo {author} {\bibfnamefont {Y.~N.}\ \bibnamefont {Joglekar}},\ and\
  \bibinfo {author} {\bibfnamefont {K.~W.}\ \bibnamefont {Murch}},\ }\bibfield
  {title} {\bibinfo {title} {Quantum state tomography across the exceptional
  point in a single dissipative qubit},\ }\href
  {https://doi.org/10.1038/s41567-019-0652-z} {\bibfield  {journal} {\bibinfo
  {journal} {Nature Physics}\ }\textbf {\bibinfo {volume} {15}},\ \bibinfo
  {pages} {1232} (\bibinfo {year} {2019})}\BibitemShut {NoStop}%
\bibitem [{\citenamefont {Sherman}\ \emph {et~al.}(2013)\citenamefont
  {Sherman}, \citenamefont {Curtis}, \citenamefont {Szwer}, \citenamefont
  {Allcock}, \citenamefont {Imreh}, \citenamefont {Lucas},\ and\ \citenamefont
  {Steane}}]{Sherman2013}%
  \BibitemOpen
  \bibfield  {author} {\bibinfo {author} {\bibfnamefont {J.~A.}\ \bibnamefont
  {Sherman}}, \bibinfo {author} {\bibfnamefont {M.~J.}\ \bibnamefont {Curtis}},
  \bibinfo {author} {\bibfnamefont {D.~J.}\ \bibnamefont {Szwer}}, \bibinfo
  {author} {\bibfnamefont {D.~T.~C.}\ \bibnamefont {Allcock}}, \bibinfo
  {author} {\bibfnamefont {G.}~\bibnamefont {Imreh}}, \bibinfo {author}
  {\bibfnamefont {D.~M.}\ \bibnamefont {Lucas}},\ and\ \bibinfo {author}
  {\bibfnamefont {A.~M.}\ \bibnamefont {Steane}},\ }\bibfield  {title}
  {\bibinfo {title} {Experimental recovery of a qubit from partial collapse},\
  }\href {https://doi.org/10.1103/PhysRevLett.111.180501} {\bibfield  {journal}
  {\bibinfo  {journal} {Phys. Rev. Lett.}\ }\textbf {\bibinfo {volume} {111}},\
  \bibinfo {pages} {180501} (\bibinfo {year} {2013})}\BibitemShut {NoStop}%
\bibitem [{\citenamefont {Brody}\ and\ \citenamefont
  {Graefe}(2012)}]{Brody2012}%
  \BibitemOpen
  \bibfield  {author} {\bibinfo {author} {\bibfnamefont {D.~C.}\ \bibnamefont
  {Brody}}\ and\ \bibinfo {author} {\bibfnamefont {E.-M.}\ \bibnamefont
  {Graefe}},\ }\bibfield  {title} {\bibinfo {title} {Mixed-state evolution in
  the presence of gain and loss},\ }\href
  {https://doi.org/10.1103/PhysRevLett.109.230405} {\bibfield  {journal}
  {\bibinfo  {journal} {Phys. Rev. Lett.}\ }\textbf {\bibinfo {volume} {109}},\
  \bibinfo {pages} {230405} (\bibinfo {year} {2012})}\BibitemShut {NoStop}%
\bibitem [{\citenamefont {Varma}\ \emph {et~al.}(2022)\citenamefont {Varma},
  \citenamefont {Muldoon}, \citenamefont {Paul}, \citenamefont {Joglekar},\
  and\ \citenamefont {Das}}]{Varma2022}%
  \BibitemOpen
  \bibfield  {author} {\bibinfo {author} {\bibfnamefont {A.~V.}\ \bibnamefont
  {Varma}}, \bibinfo {author} {\bibfnamefont {J.~E.}\ \bibnamefont {Muldoon}},
  \bibinfo {author} {\bibfnamefont {S.}~\bibnamefont {Paul}}, \bibinfo {author}
  {\bibfnamefont {Y.~N.}\ \bibnamefont {Joglekar}},\ and\ \bibinfo {author}
  {\bibfnamefont {S.}~\bibnamefont {Das}},\ }\href@noop {} {\bibinfo {title}
  {Essential role of quantum speed limit in violation of leggett-garg
  inequality across a pt-transition}} (\bibinfo {year} {2022}),\ \Eprint
  {https://arxiv.org/abs/2203.04991} {arXiv:2203.04991 [quant-ph]} \BibitemShut
  {NoStop}%
\bibitem [{\citenamefont {Ding}\ \emph {et~al.}(2021)\citenamefont {Ding},
  \citenamefont {Shi}, \citenamefont {Zhang}, \citenamefont {Shen},
  \citenamefont {Zhang},\ and\ \citenamefont {Zhang}}]{Ding2021}%
  \BibitemOpen
  \bibfield  {author} {\bibinfo {author} {\bibfnamefont {L.}~\bibnamefont
  {Ding}}, \bibinfo {author} {\bibfnamefont {K.}~\bibnamefont {Shi}}, \bibinfo
  {author} {\bibfnamefont {Q.}~\bibnamefont {Zhang}}, \bibinfo {author}
  {\bibfnamefont {D.}~\bibnamefont {Shen}}, \bibinfo {author} {\bibfnamefont
  {X.}~\bibnamefont {Zhang}},\ and\ \bibinfo {author} {\bibfnamefont
  {W.}~\bibnamefont {Zhang}},\ }\bibfield  {title} {\bibinfo {title}
  {Experimental determination of $\mathcal{P}\mathcal{T}$-symmetric exceptional
  points in a single trapped ion},\ }\href
  {https://doi.org/10.1103/PhysRevLett.126.083604} {\bibfield  {journal}
  {\bibinfo  {journal} {Phys. Rev. Lett.}\ }\textbf {\bibinfo {volume} {126}},\
  \bibinfo {pages} {083604} (\bibinfo {year} {2021})}\BibitemShut {NoStop}%
\bibitem [{\citenamefont {Miri}\ and\ \citenamefont
  {Al{\`{u}}}(2019)}]{Miri2019}%
  \BibitemOpen
  \bibfield  {author} {\bibinfo {author} {\bibfnamefont {M.-A.}\ \bibnamefont
  {Miri}}\ and\ \bibinfo {author} {\bibfnamefont {A.}~\bibnamefont
  {Al{\`{u}}}},\ }\bibfield  {title} {\bibinfo {title} {Exceptional points in
  optics and photonics},\ }\href {https://doi.org/10.1126/science.aar7709}
  {\bibfield  {journal} {\bibinfo  {journal} {Science}\ }\textbf {\bibinfo
  {volume} {363}},\ \bibinfo {pages} {eaar7709} (\bibinfo {year}
  {2019})}\BibitemShut {NoStop}%
\bibitem [{\citenamefont {Hodaei}\ \emph {et~al.}(2017)\citenamefont {Hodaei},
  \citenamefont {Hassan}, \citenamefont {Wittek}, \citenamefont
  {Garcia-Gracia}, \citenamefont {El-Ganainy}, \citenamefont
  {Christodoulides},\ and\ \citenamefont {Khajavikhan}}]{Hodaei2017}%
  \BibitemOpen
  \bibfield  {author} {\bibinfo {author} {\bibfnamefont {H.}~\bibnamefont
  {Hodaei}}, \bibinfo {author} {\bibfnamefont {A.~U.}\ \bibnamefont {Hassan}},
  \bibinfo {author} {\bibfnamefont {S.}~\bibnamefont {Wittek}}, \bibinfo
  {author} {\bibfnamefont {H.}~\bibnamefont {Garcia-Gracia}}, \bibinfo {author}
  {\bibfnamefont {R.}~\bibnamefont {El-Ganainy}}, \bibinfo {author}
  {\bibfnamefont {D.~N.}\ \bibnamefont {Christodoulides}},\ and\ \bibinfo
  {author} {\bibfnamefont {M.}~\bibnamefont {Khajavikhan}},\ }\bibfield
  {title} {\bibinfo {title} {Enhanced sensitivity at higher-order exceptional
  points},\ }\href {https://doi.org/10.1038/nature23280} {\bibfield  {journal}
  {\bibinfo  {journal} {Nature}\ }\textbf {\bibinfo {volume} {548}},\ \bibinfo
  {pages} {187} (\bibinfo {year} {2017})}\BibitemShut {NoStop}%
\bibitem [{\citenamefont {Chen}\ \emph {et~al.}(2017)\citenamefont {Chen},
  \citenamefont {\"{O}zdemir}, \citenamefont {Zhao}, \citenamefont {Wiersig},\
  and\ \citenamefont {Yang}}]{Chen2017}%
  \BibitemOpen
  \bibfield  {author} {\bibinfo {author} {\bibfnamefont {W.}~\bibnamefont
  {Chen}}, \bibinfo {author} {\bibfnamefont {{\c{S}}.~K.}\ \bibnamefont
  {\"{O}zdemir}}, \bibinfo {author} {\bibfnamefont {G.}~\bibnamefont {Zhao}},
  \bibinfo {author} {\bibfnamefont {J.}~\bibnamefont {Wiersig}},\ and\ \bibinfo
  {author} {\bibfnamefont {L.}~\bibnamefont {Yang}},\ }\bibfield  {title}
  {\bibinfo {title} {Exceptional points enhance sensing in an optical
  microcavity},\ }\href {https://doi.org/10.1038/nature23281} {\bibfield
  {journal} {\bibinfo  {journal} {Nature}\ }\textbf {\bibinfo {volume} {548}},\
  \bibinfo {pages} {192} (\bibinfo {year} {2017})}\BibitemShut {NoStop}%
\bibitem [{\citenamefont {Doppler}\ \emph {et~al.}(2016)\citenamefont
  {Doppler}, \citenamefont {Mailybaev}, \citenamefont {B\"{o}hm}, \citenamefont
  {Kuhl}, \citenamefont {Girschik}, \citenamefont {Libisch}, \citenamefont
  {Milburn}, \citenamefont {Rabl}, \citenamefont {Moiseyev},\ and\
  \citenamefont {Rotter}}]{Doppler2016}%
  \BibitemOpen
  \bibfield  {author} {\bibinfo {author} {\bibfnamefont {J.}~\bibnamefont
  {Doppler}}, \bibinfo {author} {\bibfnamefont {A.~A.}\ \bibnamefont
  {Mailybaev}}, \bibinfo {author} {\bibfnamefont {J.}~\bibnamefont {B\"{o}hm}},
  \bibinfo {author} {\bibfnamefont {U.}~\bibnamefont {Kuhl}}, \bibinfo {author}
  {\bibfnamefont {A.}~\bibnamefont {Girschik}}, \bibinfo {author}
  {\bibfnamefont {F.}~\bibnamefont {Libisch}}, \bibinfo {author} {\bibfnamefont
  {T.~J.}\ \bibnamefont {Milburn}}, \bibinfo {author} {\bibfnamefont
  {P.}~\bibnamefont {Rabl}}, \bibinfo {author} {\bibfnamefont {N.}~\bibnamefont
  {Moiseyev}},\ and\ \bibinfo {author} {\bibfnamefont {S.}~\bibnamefont
  {Rotter}},\ }\bibfield  {title} {\bibinfo {title} {Dynamically encircling an
  exceptional point for asymmetric mode switching},\ }\href
  {https://doi.org/10.1038/nature18605} {\bibfield  {journal} {\bibinfo
  {journal} {Nature}\ }\textbf {\bibinfo {volume} {537}},\ \bibinfo {pages}
  {76} (\bibinfo {year} {2016})}\BibitemShut {NoStop}%
\bibitem [{\citenamefont {Xu}\ \emph {et~al.}(2016)\citenamefont {Xu},
  \citenamefont {Mason}, \citenamefont {Jiang},\ and\ \citenamefont
  {Harris}}]{Xu2016}%
  \BibitemOpen
  \bibfield  {author} {\bibinfo {author} {\bibfnamefont {H.}~\bibnamefont
  {Xu}}, \bibinfo {author} {\bibfnamefont {D.}~\bibnamefont {Mason}}, \bibinfo
  {author} {\bibfnamefont {L.}~\bibnamefont {Jiang}},\ and\ \bibinfo {author}
  {\bibfnamefont {J.~G.~E.}\ \bibnamefont {Harris}},\ }\bibfield  {title}
  {\bibinfo {title} {Topological energy transfer in an optomechanical system
  with exceptional points},\ }\href {https://doi.org/10.1038/nature18604}
  {\bibfield  {journal} {\bibinfo  {journal} {Nature}\ }\textbf {\bibinfo
  {volume} {537}},\ \bibinfo {pages} {80} (\bibinfo {year} {2016})}\BibitemShut
  {NoStop}%
\bibitem [{\citenamefont {Wang}\ \emph {et~al.}(2021)\citenamefont {Wang},
  \citenamefont {Dutt}, \citenamefont {Wojcik},\ and\ \citenamefont
  {Fan}}]{Wang2021}%
  \BibitemOpen
  \bibfield  {author} {\bibinfo {author} {\bibfnamefont {K.}~\bibnamefont
  {Wang}}, \bibinfo {author} {\bibfnamefont {A.}~\bibnamefont {Dutt}}, \bibinfo
  {author} {\bibfnamefont {C.~C.}\ \bibnamefont {Wojcik}},\ and\ \bibinfo
  {author} {\bibfnamefont {S.}~\bibnamefont {Fan}},\ }\bibfield  {title}
  {\bibinfo {title} {Topological complex-energy braiding of non-hermitian
  bands},\ }\href {https://doi.org/10.1038/s41586-021-03848-x} {\bibfield
  {journal} {\bibinfo  {journal} {Nature}\ }\textbf {\bibinfo {volume} {598}},\
  \bibinfo {pages} {59} (\bibinfo {year} {2021})}\BibitemShut {NoStop}%
\bibitem [{\citenamefont {Patil}\ \emph {et~al.}(2022)\citenamefont {Patil},
  \citenamefont {H\"{o}ller}, \citenamefont {Henry}, \citenamefont {Guria},
  \citenamefont {Zhang}, \citenamefont {Jiang}, \citenamefont {Kralj},
  \citenamefont {Read},\ and\ \citenamefont {Harris}}]{Patil2022}%
  \BibitemOpen
  \bibfield  {author} {\bibinfo {author} {\bibfnamefont {Y.~S.~S.}\
  \bibnamefont {Patil}}, \bibinfo {author} {\bibfnamefont {J.}~\bibnamefont
  {H\"{o}ller}}, \bibinfo {author} {\bibfnamefont {P.~A.}\ \bibnamefont
  {Henry}}, \bibinfo {author} {\bibfnamefont {C.}~\bibnamefont {Guria}},
  \bibinfo {author} {\bibfnamefont {Y.}~\bibnamefont {Zhang}}, \bibinfo
  {author} {\bibfnamefont {L.}~\bibnamefont {Jiang}}, \bibinfo {author}
  {\bibfnamefont {N.}~\bibnamefont {Kralj}}, \bibinfo {author} {\bibfnamefont
  {N.}~\bibnamefont {Read}},\ and\ \bibinfo {author} {\bibfnamefont {J.~G.~E.}\
  \bibnamefont {Harris}},\ }\bibfield  {title} {\bibinfo {title} {Measuring the
  knot of non-hermitian degeneracies and non-commuting braids},\ }\href
  {https://doi.org/10.1038/s41586-022-04796-w} {\bibfield  {journal} {\bibinfo
  {journal} {Nature}\ }\textbf {\bibinfo {volume} {607}},\ \bibinfo {pages}
  {271} (\bibinfo {year} {2022})}\BibitemShut {NoStop}%
\bibitem [{\citenamefont {Abbasi}\ \emph {et~al.}(2022)\citenamefont {Abbasi},
  \citenamefont {Chen}, \citenamefont {Naghiloo}, \citenamefont {Joglekar},\
  and\ \citenamefont {Murch}}]{Abbasi2022}%
  \BibitemOpen
  \bibfield  {author} {\bibinfo {author} {\bibfnamefont {M.}~\bibnamefont
  {Abbasi}}, \bibinfo {author} {\bibfnamefont {W.}~\bibnamefont {Chen}},
  \bibinfo {author} {\bibfnamefont {M.}~\bibnamefont {Naghiloo}}, \bibinfo
  {author} {\bibfnamefont {Y.~N.}\ \bibnamefont {Joglekar}},\ and\ \bibinfo
  {author} {\bibfnamefont {K.~W.}\ \bibnamefont {Murch}},\ }\bibfield  {title}
  {\bibinfo {title} {Topological quantum state control through
  exceptional-point proximity},\ }\href
  {https://doi.org/10.1103/PhysRevLett.128.160401} {\bibfield  {journal}
  {\bibinfo  {journal} {Phys. Rev. Lett.}\ }\textbf {\bibinfo {volume} {128}},\
  \bibinfo {pages} {160401} (\bibinfo {year} {2022})}\BibitemShut {NoStop}%
\bibitem [{\citenamefont {Wu}\ \emph {et~al.}(2019)\citenamefont {Wu},
  \citenamefont {Liu}, \citenamefont {Geng}, \citenamefont {Song},
  \citenamefont {Ye}, \citenamefont {Duan}, \citenamefont {Rong},\ and\
  \citenamefont {Du}}]{Wu2019}%
  \BibitemOpen
  \bibfield  {author} {\bibinfo {author} {\bibfnamefont {Y.}~\bibnamefont
  {Wu}}, \bibinfo {author} {\bibfnamefont {W.}~\bibnamefont {Liu}}, \bibinfo
  {author} {\bibfnamefont {J.}~\bibnamefont {Geng}}, \bibinfo {author}
  {\bibfnamefont {X.}~\bibnamefont {Song}}, \bibinfo {author} {\bibfnamefont
  {X.}~\bibnamefont {Ye}}, \bibinfo {author} {\bibfnamefont {C.-K.}\
  \bibnamefont {Duan}}, \bibinfo {author} {\bibfnamefont {X.}~\bibnamefont
  {Rong}},\ and\ \bibinfo {author} {\bibfnamefont {J.}~\bibnamefont {Du}},\
  }\bibfield  {title} {\bibinfo {title} {Observation of parity-time symmetry
  breaking in a single-spin system},\ }\href
  {https://doi.org/10.1126/science.aaw8205} {\bibfield  {journal} {\bibinfo
  {journal} {Science}\ }\textbf {\bibinfo {volume} {364}},\ \bibinfo {pages}
  {878} (\bibinfo {year} {2019})}\BibitemShut {NoStop}%
\bibitem [{\citenamefont {Liu}\ \emph {et~al.}(2021)\citenamefont {Liu},
  \citenamefont {Wu}, \citenamefont {Duan}, \citenamefont {Rong},\ and\
  \citenamefont {Du}}]{Liu2021}%
  \BibitemOpen
  \bibfield  {author} {\bibinfo {author} {\bibfnamefont {W.}~\bibnamefont
  {Liu}}, \bibinfo {author} {\bibfnamefont {Y.}~\bibnamefont {Wu}}, \bibinfo
  {author} {\bibfnamefont {C.-K.}\ \bibnamefont {Duan}}, \bibinfo {author}
  {\bibfnamefont {X.}~\bibnamefont {Rong}},\ and\ \bibinfo {author}
  {\bibfnamefont {J.}~\bibnamefont {Du}},\ }\bibfield  {title} {\bibinfo
  {title} {Dynamically encircling an exceptional point in a real quantum
  system},\ }\href {https://doi.org/10.1103/PhysRevLett.126.170506} {\bibfield
  {journal} {\bibinfo  {journal} {Phys. Rev. Lett.}\ }\textbf {\bibinfo
  {volume} {126}},\ \bibinfo {pages} {170506} (\bibinfo {year}
  {2021})}\BibitemShut {NoStop}%
\bibitem [{\citenamefont {Maraviglia}\ \emph {et~al.}(2022)\citenamefont
  {Maraviglia}, \citenamefont {Yard}, \citenamefont {Wakefield}, \citenamefont
  {Carolan}, \citenamefont {Sparrow}, \citenamefont {Chakhmakhchyan},
  \citenamefont {Harrold}, \citenamefont {Hashimoto}, \citenamefont {Matsuda},
  \citenamefont {Harter}, \citenamefont {Joglekar},\ and\ \citenamefont
  {Laing}}]{Maraviglia2022}%
  \BibitemOpen
  \bibfield  {author} {\bibinfo {author} {\bibfnamefont {N.}~\bibnamefont
  {Maraviglia}}, \bibinfo {author} {\bibfnamefont {P.}~\bibnamefont {Yard}},
  \bibinfo {author} {\bibfnamefont {R.}~\bibnamefont {Wakefield}}, \bibinfo
  {author} {\bibfnamefont {J.}~\bibnamefont {Carolan}}, \bibinfo {author}
  {\bibfnamefont {C.}~\bibnamefont {Sparrow}}, \bibinfo {author} {\bibfnamefont
  {L.}~\bibnamefont {Chakhmakhchyan}}, \bibinfo {author} {\bibfnamefont
  {C.}~\bibnamefont {Harrold}}, \bibinfo {author} {\bibfnamefont
  {T.}~\bibnamefont {Hashimoto}}, \bibinfo {author} {\bibfnamefont
  {N.}~\bibnamefont {Matsuda}}, \bibinfo {author} {\bibfnamefont {A.~K.}\
  \bibnamefont {Harter}}, \bibinfo {author} {\bibfnamefont {Y.~N.}\
  \bibnamefont {Joglekar}},\ and\ \bibinfo {author} {\bibfnamefont
  {A.}~\bibnamefont {Laing}},\ }\bibfield  {title} {\bibinfo {title} {Photonic
  quantum simulations of coupled $\mathcal{PT}$-symmetric hamiltonians},\
  }\href {https://doi.org/10.1103/PhysRevResearch.4.013051} {\bibfield
  {journal} {\bibinfo  {journal} {Phys. Rev. Res.}\ }\textbf {\bibinfo {volume}
  {4}},\ \bibinfo {pages} {013051} (\bibinfo {year} {2022})}\BibitemShut
  {NoStop}%
\bibitem [{\citenamefont {Gerritsma}\ \emph {et~al.}(2008)\citenamefont
  {Gerritsma}, \citenamefont {Kirchmair}, \citenamefont {Zähringer},
  \citenamefont {Benhelm}, \citenamefont {Blatt},\ and\ \citenamefont
  {Roos}}]{Gerritsma2008}%
  \BibitemOpen
  \bibfield  {author} {\bibinfo {author} {\bibfnamefont {R.}~\bibnamefont
  {Gerritsma}}, \bibinfo {author} {\bibfnamefont {G.}~\bibnamefont
  {Kirchmair}}, \bibinfo {author} {\bibfnamefont {F.}~\bibnamefont
  {Zähringer}}, \bibinfo {author} {\bibfnamefont {J.}~\bibnamefont {Benhelm}},
  \bibinfo {author} {\bibfnamefont {R.}~\bibnamefont {Blatt}},\ and\ \bibinfo
  {author} {\bibfnamefont {C.~F.}\ \bibnamefont {Roos}},\ }\bibfield  {title}
  {\bibinfo {title} {Precision measurement of the branching fractions of the
  4p 2{P}3/2 decay of {C}a {I}{I}},\ }\href
  {https://doi.org/10.1140/epjd/e2008-00196-9} {\bibfield  {journal} {\bibinfo
  {journal} {The European Physical Journal D}\ }\textbf {\bibinfo {volume}
  {50}},\ \bibinfo {pages} {13} (\bibinfo {year} {2008})}\BibitemShut {NoStop}%
\bibitem [{\citenamefont {Brion}\ \emph {et~al.}(2007)\citenamefont {Brion},
  \citenamefont {Pedersen},\ and\ \citenamefont {M{\o}lmer}}]{Brion2007}%
  \BibitemOpen
  \bibfield  {author} {\bibinfo {author} {\bibfnamefont {E.}~\bibnamefont
  {Brion}}, \bibinfo {author} {\bibfnamefont {L.~H.}\ \bibnamefont
  {Pedersen}},\ and\ \bibinfo {author} {\bibfnamefont {K.}~\bibnamefont
  {M{\o}lmer}},\ }\bibfield  {title} {\bibinfo {title} {Adiabatic elimination
  in a lambda system},\ }\href {https://doi.org/10.1088/1751-8113/40/5/011}
  {\bibfield  {journal} {\bibinfo  {journal} {Journal of Physics A:
  Mathematical and Theoretical}\ }\textbf {\bibinfo {volume} {40}},\ \bibinfo
  {pages} {1033} (\bibinfo {year} {2007})}\BibitemShut {NoStop}%
\end{thebibliography}%



\clearpage
\pagebreak
\setcounter{equation}{0}
\setcounter{figure}{0}
\setcounter{table}{0}
\setcounter{page}{1}
\makeatletter
\renewcommand{\theequation}{A\arabic{equation}}
\renewcommand{\thefigure}{S\arabic{figure}}
\renewcommand{\citenumfont}[1]{S#1}
\makeatother

\begin{widetext}
\section*{Methods}

\begin{figure*}
\centering
\includegraphics[width=\textwidth]{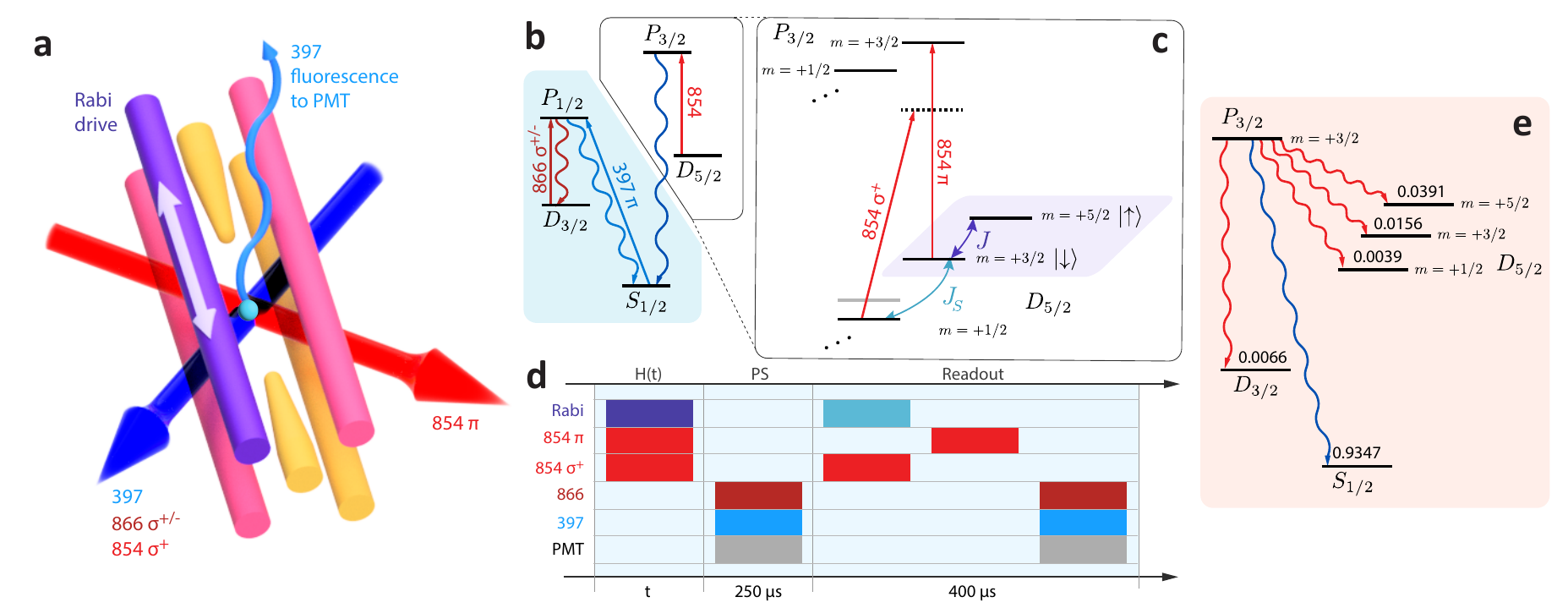}
\caption{{\bf Experiment setup and procedure.} {\bf a.} Schematic render of our trap (\textit{pink}---RF quadrupole electrodes, \textit{yellow, purple}---DC electrodes) and relevant control fields, with lavender arrow representing the AC current supplying the magnetic qubit drive and cyan dot representing a $^{40}$Ca$^+$ ion. {\bf b.} Electronic energy level diagram for $^{40}$Ca$^+$ showing wavelengths associated with relevant transitions.  The set of states involved in the $S_{1/2}$ to $P_{1/2}$ cycling transition used for Doppler cooling and state detection are highlighted in blue. {\bf c.} Zoom-in on the $P_{3/2}$ and $D_{5/2}$ levels, showing the spin states in which the qubit is encoded (highlighted in purple) and the spin states and control fields relevant to the qubit drive, including the RF Rabi drive between the qubit states ($J$, purple) and the RF shelving drive for state readout ($J_S$, turquoise). {\bf d.} Experiment pulse sequence, not including Doppler cooling and state initialization. Approximate times for post-selection (PS) and readout are shown below. {\bf e.} Branching ratios from the $\ket{ m= +3/2}$ state of the $P_{3/2}$ manifold. }
\label{fig:four}
\end{figure*}
\end{widetext}

\subsection{Experiment setup}
\label{subsec:expt}

We realize a non-Hermitian Hamiltonian by using a set of control fields to drive a single $^{40}$Ca$^+$ ion trapped in a macroscopic (0.75 mm ion-electrode spacing), room-temperature, linear-Paul trap (all represented schematically in Fig.~\ref{fig:four}a).  Following Sherman et al.~\cite{Sherman2013}, we choose qubit states in the metastable $D_{5/2}$ manifold and employ a scheme for post-selecting on decay/deshelving from these states. A summary of the electronic states and associated control fields relevant to qubit control, post-selection, and readout are shown in Fig.~\ref{fig:four}b and Fig.~\ref{fig:four}c.

\noindent{\bf Qubit states and preparation:} We use as qubit states $\ket{\uparrow} \equiv \ket{m= + 5/2}$ and $\ket{\downarrow} \equiv\ket{ m= +3/2}$ within the metastable $D_{5/2}$ manifold, where $m$ is the total angular momentum projection along the direction of the quantization magnetic field of approximately 0.498\,G.  This magnetic field gives a qubit frequency $\omega_0 \approx 2\pi\times 2.63$\,MHz. To isolate the qubit states from the rest of the $D_{5/2}$ manifold, we apply a far-detuned, high-intensity $\sigma^+$-polarized 854\,nm laser beam which induces an AC Stark shift on states outside of the qubit, including an $\approx -2\pi\times 0.5$\,MHz shift on the $\ket{ m= +1/2}$ level.  We initialize the qubit in $\ket{\uparrow}$ through optical pumping from the $S_{1/2}$ ground state after performing Doppler cooling.

 \noindent{\bf Generating a non-Hermitian Hamiltonian:} We use two control fields to generate the non-Hermitian Hamiltonian,
\begin{equation}
    \hat{H} = J\sigma_x + i\gamma\sigma_z.
\end{equation} 
The $\sigma_x$ coupling is performed using resonant RF pulses at the qubit frequency. Realizing the $\sigma_z$ type term requires coupling only the $\ket{\downarrow}$ state to the short lived $P_{3/2}$ manifold using a $\pi$-polarized, near-resonant 854\,nm laser beam, where population then primarily decays to the $S_{1/2}$ state.  These processes are illustrated in Fig.~\ref{fig:four}c.  It should be noted that the qubit frequency must be recalibrated at each intensity of the 854\,nm beam used, since this beam induces a light shift on the $\ket{\downarrow}$ state. 

\noindent{\bf Post-selection:} After applying the non-Hermitian Hamiltonian described above, we post-select on decay from $P_{3/2}$.  To do this, we look for population in the $S_{1/2}$ and $D_{3/2}$ states by simultaneously driving the 397\,nm $S_{1/2}$ to $P_{1/2}$ cycling transition and the 866\,nm $D_{3/2}$ to $P_{1/2}$ transition, rejecting runs of the experiment where 397\,nm fluorescence is detected. An important limitation of this post-selection scheme is that it only detects when the ion has decayed from $P_{3/2}$ to $S_{1/2}$ or $D_{3/2}$, when 5.87\% of the $P_{3/2}$ population decays back to the $D_{5/2}$ state, with the exact branching ratios to the different $D_{5/2}$ spin states shown in Fig.~\ref{fig:four}d~\cite{Gerritsma2008}.  The inability to detect these decay events functionally limits our evolution time.

 \noindent{\bf State readout:} If a run of the experiment is not rejected in post-selection, we read out by deshelving population in the $\ket{\downarrow}$ state using a $\pi$-polarized 854\,nm laser beam.  To reduce the chance of decay to the $\ket{\uparrow}$ state during this deshelving process, which leads to a readout error, we first drive population from the $\ket{\downarrow}$ to the $\ket{ m = +1/2}$ state using a resonant RF $\pi$ pulse at a frequency $\approx 3.1$\,MHz.  We then repeat the process used for post-selection, driving the $S_{1/2}$ to $P_{1/2}$ and $D_{3/2}$ to $P_{1/2}$ transitions and looking for fluorescence.

 The full experimental sequence, from the application of the non-Hermitian Hamiltonian to state readout, is represented in Fig.~\ref{fig:four}e.


\subsection{From 4-level Lindblad to a 2-level non-Hermitian Hamiltonian by post-selection}
\label{subsec:lind}
The decohering dynamics of the single $^{40}$Ca$^{+}$ ion are described by the Gorini-Kossakowski–Sudarshan–Lindblad equation, 
\begin{align}
\partial_{t}\rho(t)&= -\imath[H_{0},\rho] + \sum_{\mu}\mathcal{D}_{\mu}[\rho]\equiv\mathcal{L}[\rho] \\
    \mathcal{D}_{\alpha}[\rho] &= L_{\alpha}\rho L_{\alpha}^{\dagger} - \frac{1}{2}\left(L_{\alpha}^{\dagger}L_{\alpha}\rho + \rho L_{\alpha}^{\dagger}L_{\alpha}\right)
\end{align}
where $\{L_{\alpha}\}$ are Lindblad dissipators that characterize the coupling of the ion to the environment. In our model, the ion has 4-levels $\{\ket{\uparrow},\ket{\downarrow},\ket{A},\ket{g}\}$, with two coherent drives, 
\begin{align}
    H_{0} &= J(\ket{\uparrow}\bra{\downarrow}+\ket{\downarrow}\bra{\uparrow})+ J_{A}(\ket{\downarrow}\bra{A}+\ket{A}\bra{\downarrow}).
\end{align}
and three most relevant spontaneous-emission dissipators 
\begin{align}
    L_{\uparrow} &= \sqrt{\gamma_{\uparrow}}\ket{\uparrow}\bra{A}, \\
    L_{\downarrow} &= \sqrt{\gamma_{\downarrow}}\ket{\downarrow}\bra{A}, \\
    L_{g} &= \sqrt{\gamma_{g}}\ket{g}\bra{A},
\end{align}
that represent the decay from the $P_{3/2}$ auxiliary level $\ket{A}$ to the $D_{5/2}$ qubit manifold $\{\ket{\uparrow},\ket{\downarrow}\}$ and the $S_{1/2}$ ground state $\ket{g}$. Since $\gamma_g\gg\{J_A,\gamma_{\uparrow},\gamma_{\downarrow}\}$, we only consider the dissipator $L_g$ analytically and treat other dissipators as sources of errors (such as backflow to the qubit manifold) that become relevant at long times.

We analytically diagonalize the $16\times 16$ vectorized superoperator $\mathcal{L}_4$, approximate its eigenvalues and eigenvectors under the constraint $\gamma_g\gg J_A$, and compare the resulting approximate Lindbladian $\mathcal{L}_{4a}$ to the $9\times 9$ vectorized Lindbladian $\mathcal{L}_3$ of a 3-level system $\{\ket{\uparrow},\ket{\downarrow},\ket{g}\}$ with qubit-manifold Rabi drive $J$ and a single dissipator $L_\textrm{eff}=\sqrt{\gamma}\ket{g}\bra{\downarrow}$. By adiabatic elimination of the $\ket{A}$ state~\cite{Brion2007} $\mathcal{L}_{4a}$ maps onto $\mathcal{L}_3$ with $\gamma=J_A^2/\gamma_g$. As a result, post-selection over no-quantum-jumps to the $\ket{g}$ state trajectories yields a 2-level system with Hamiltonian
\begin{align}
H_{\uparrow\downarrow} &\approx J\sigma_{x}+i\left(\frac{J_{A}^{2}}{\gamma_{g}}\right)\sigma_z. 
\end{align}


\subsection{Optimizing $K_{3}(\gamma,t)$ and transit time $T_{\downarrow\rightarrow\uparrow}(\gamma)$}

\begin{figure*}
\centering
\includegraphics[width=\textwidth]{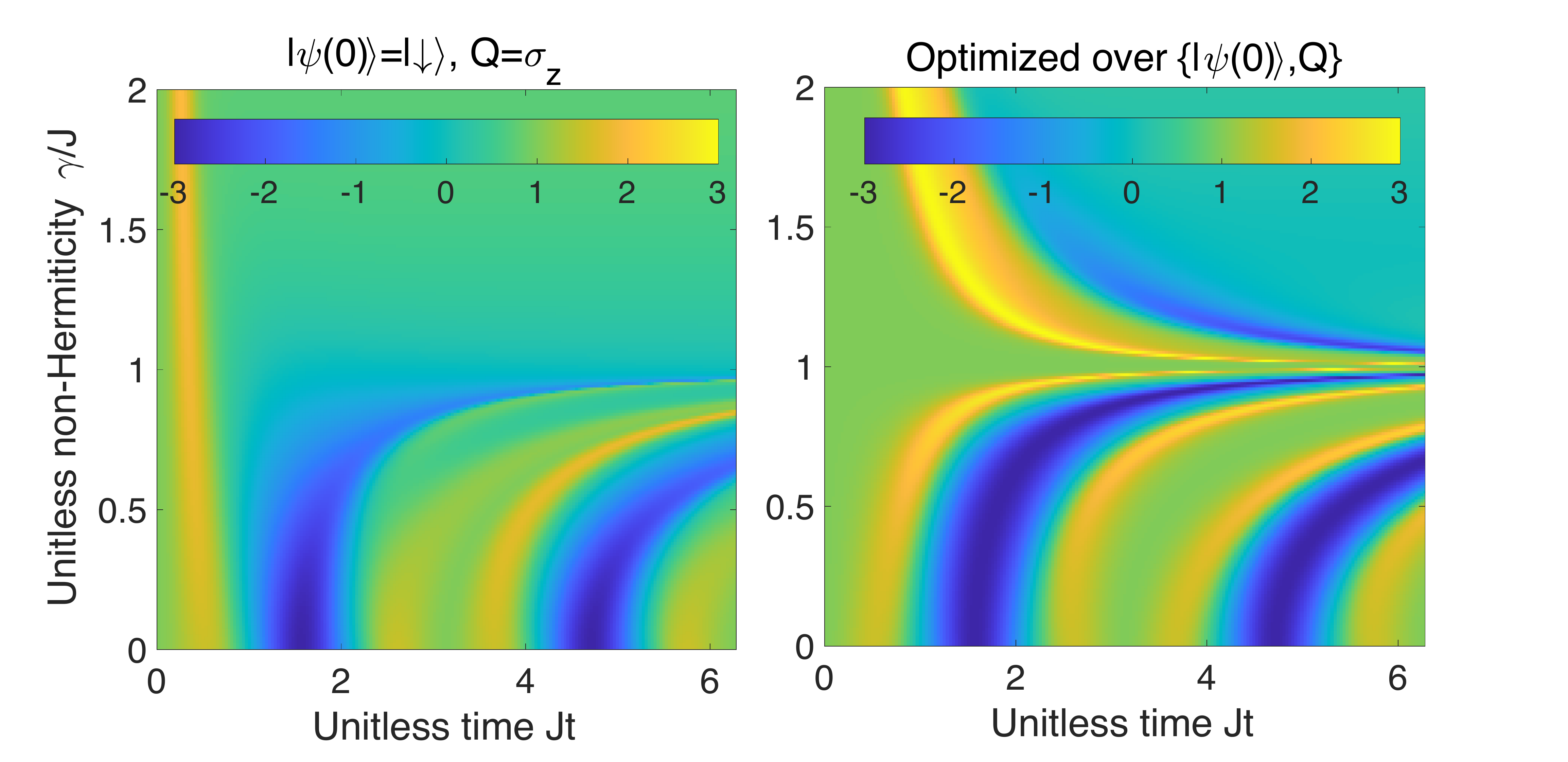}
\caption{(a) Leggett-Garg parameter $K_{3}$ with initial state $\ket{\psi(0)}=\ket{\downarrow}$ and projectively measured observable $Q=\sigma_{z}$ shows a maximum exceeding the L\"{u}der bound occurring at times $Jt\leq \pi/6\approx 0.523$. (b) Maximum $K_{3}(\gamma,t)$ possible when both $\ket{\psi(0)}$ and $Q$ are allowed to vary shows a broad region with $K_3>1.5$ past the EP at $\gamma/J=1$~\cite{Varma2022}. In contrast to (a), however, this region occurs at times that are longer than the unitary-limit value, $Jt>\pi/6$. Thus, with this protocol, stronger temporal correlations develop, but slowly.}
\label{fig:Optimized}
\end{figure*}

For equally spaced times $t_1=0,t_2=t,t_3=2t$, the LG parameter $K_3(t)$ is given by 
\begin{align}
\label{eq:k3m}
K_3(t)& = C(t)+F(t)-C(2t),
\end{align}
where $C_{32}(t)\equiv F(t)\neq C_{21}(t)\equiv C(t)$ due to the norm-preserving, state-dependent, nonlinear term in the equation of motion in the post-selected qubit manifold~\cite{Brody2012,Varma2022}. For initial state $\ket{\psi(0)}=\ket{\downarrow}$ the correlation functions $C(t)$ and $F(t)$ are given by
\begin{widetext}
\begin{align}
\label{eq:cm}
C(t)&=\frac{\left[\cos(\Delta t)-\gamma\sin(\Delta t)/\Delta)\right]^2-J^2\sin^2(\Delta t)/\Delta^2}{\left[\cos(\Delta t)-\gamma\sin(\Delta t)/\Delta\right]^2+J^2\sin^2(\Delta t)/\Delta^2}.\\
\label{eq:fm}
F(t)&=\frac{\left[\cos(\Delta t)-\gamma\sin(\Delta t)/\Delta\right]^2 C(t)+\left[J^2\sin^2(\Delta t)/\Delta^2\right] C(-t)}{\left[\cos(\Delta t)-\gamma\sin(\Delta t)/\Delta\right]^2+J^2\sin^2(\Delta t)/\Delta^2}.
\end{align}
\end{widetext}
The EP limit in Eqs.(\ref{eq:cm})-(\ref{eq:fm}) is obtained by $\Delta\rightarrow 0$ and using $\sin(\Delta t)/\Delta\rightarrow t$. Figure~\ref{fig:Optimized}a shows $K_3$ as a function of dimensionless time $Jt$ and dimensionless non-Hermiticity $\gamma/J$, with the peak moving to higher values at smaller times. In contrast, Fig.~\ref{fig:Optimized}b shows optimized $K_3$ where the initial state $\ket{\psi(0)}$ and the dichotomous observable $Q$ are allowed to vary~\cite{Varma2022}. 

The transit time $T_{\downarrow\rightarrow\uparrow}(\gamma)$ is determined from $\ket{\psi(t)}=G(t)\ket{\downarrow}/\sqrt{\bra{\downarrow}G^\dagger(t)G(t)\ket{\downarrow}}$ by requiring that its amplitude on the $\ket{\downarrow}$ basis-state be zero, i.e. 
$\cos(\Delta t)=\gamma\sin(\Delta t)/\Delta$. This equation has a positive solution irrespective of whether $\Delta$ is real, zero, or purely imaginary, 
\begin{align}
T_{\downarrow\rightarrow\uparrow}(\gamma)&=\frac{1}{\Delta}\arctan(\frac{\Delta}{\gamma}). 
\end{align}
At $\gamma=0$, we get $T_{\downarrow\rightarrow\uparrow}=\tau_{\textrm{QSL}}=\pi/2J$. In the limit $\gamma/J\gg 1$, we get a small $T_{\downarrow\rightarrow\uparrow}\approx\gamma^{-1}\ln(2\gamma/J)$. In contrast, starting from state $\ket{\uparrow}$, the transit time $T_{\uparrow\rightarrow\downarrow}$ to the antipodal state $\ket{\downarrow}$ is defined by the equation 
\begin{align}
\label{eq:tdown}
    \cos(\Delta T_{\uparrow\rightarrow\downarrow})=-\frac{\gamma}{\Delta}\sin(\Delta T_{\uparrow\rightarrow\downarrow}).
\end{align}
At $\gamma=0$, we obtain the expected answer $T_{\uparrow\rightarrow\downarrow}=\tau_{\textrm{QSL}}$ as the answer. When $\gamma>J$, the  non-periodic nature of hyperbolic counterparts in Eq.(\ref{eq:tdown}) precludes any positive solution for the transit time. This is consistent with the observation that the gain-state $\ket{\uparrow}$ does not completely transit to the loss-state $\ket{\downarrow}$ in the $\mathcal{PT}$-symmetry broken region. In the intermediate region $\gamma<J$, by using the identity $\tan(\pi-x)=-\tan(x)$, we get 
\begin{align}
T_{\uparrow\rightarrow\downarrow}(\gamma)+T_{\downarrow\rightarrow\uparrow}(\gamma)=\frac{\pi}{\Delta}.
\end{align}


\end{document}